\documentclass[10pt,preprint,twocolumn]{aastex631}
\usepackage{amsmath}
\usepackage{subfigure}
\usepackage{hyperref}

\usepackage{url}
\usepackage{graphicx}
\usepackage{natbib}
\usepackage{amssymb,txfonts}
\usepackage{multirow}
\usepackage{array}
\usepackage{sidecap}
\usepackage{hyperref}
\usepackage{epstopdf}
\usepackage{footnote}
\usepackage{tabularx}
\usepackage{booktabs}
\usepackage{soul}
\newcommand{\p}[1]{{\color{red}{#1}}}

\newcommand{\speed}[1]{#1 km~s${}^{-1}$}

\newcommand{\rmnum}[1]{\uppercase\expandafter{\romannumeral #1}}
\usepackage{graphicx} \usepackage{epstopdf}
\shorttitle{jet-CME}
\shortauthors{Duan et al.}

\begin{document}

\title{On the Determining Physical Factor of Jet-Related Coronal Mass Ejection's Morphology in the High Corona}
\correspondingauthor{Yuandeng Shen}
\email{ydshen@ynao.ac.cn}

\author[0000-0001-9491-699X]{Yadan Duan}
\affiliation{School of Earth and Space Sciences, Peking University, Beijing, 100871, People's Republic of China}
\author[0000-0001-9493-4418]{Yuandeng Shen}
\affiliation{Yunnan Observatories, Chinese Academy of Sciences, Kunming, 650216, People's Republic of China}
\affiliation{Shenzhen Key Laboratory of Numerical Prediction for Space Storm, Institute of Space
Science and Applied Technology, Harbin Institute of Technology, Shenzhen 518055, People's Republic of China}
\author{Zehao Tang}
\affiliation{Yunnan Observatories, Chinese Academy of Sciences, Kunming, 650216, People's Republic of China}
\affiliation{University of Chinese Academy of Sciences, Beijing 100049, People's Republic of China}
\author{Chenrui Zhou}
\affiliation{Yunnan Observatories, Chinese Academy of Sciences, Kunming, 650216, People's Republic of China}
\affiliation{University of Chinese Academy of Sciences, Beijing 100049, People's Republic of China}
\author{Song Tan}
\affiliation{Leibniz Institute for Astrophysics Potsdam (AIP), An der Sternwarte 16, 14482 Potsdam, Germany}

\begin{abstract}
A solar jet can often cause coronal mass ejections (CMEs) with different morphologies in the high corona, for example, jet-like CMEs, bubble-like CMEs, and so-called twin CMEs that include a pair of simultaneous jet-like and bubble-like CMEs. However, what determines the morphology of a jet-related CME is still an open question. Using high spatiotemporal resolution stereoscopic observations taken by the Solar Dynamics Observatory (SDO) and the Solar Terrestrial Relations Observatory (STEREO) from October 2010 to December 2012, we performed a statistical study of jet-related CMEs to study the potential physical factors that determine the morphology of CMEs in the outer corona. Our statistical sample includes 16 jet-related CME events of which 7 are twin CME events and 9 are jet-like narrow CMEs. We find that all CMEs in our sample were accompanied by filament-driven blowout jets and Type \rmnum{3} radio bursts during their initial formation and involved magnetic reconnection between filament channels and the surrounding magnetic fields. Most of our cases occurred in a fan-spine magnetic configuration. Our study suggests that the bubble-like components of twin CMEs lacking an obvious core are related to the expansion of the closed-loop systems next to the fan-spine topology, while the jet-like component is from the coronal extension of the jet plasma along open fields. Based on the statistical results, we conclude that the morphology of jet-related CMEs in the high corona may be related to the filament length and the initial magnetic null point height of the fan-spine structures.

\end{abstract}


\section{Introduction} \label{sec:intro}
Coronal mass ejections (CMEs) are eruptions of large-scale magnetized plasma on the Sun. CMEs eject a large amount of material and magnetic flux into interplanetary space, which is the main driving force of space weather around Earth \citep{2011LRSP....8....1C}. Space weather forecasting is necessary because these materials and magnetic fluxes can even generate currents within the Earth's crust that cause severe electrical damage to installations \citep{2005AdSpR..36.2231P}. The key to space weather forecasting lies in clarifying the early evolution of CMEs, such as their initiation and dynamics processes \citep[e.g.,][]{2023JApA...44...20M}. Typically, CMEs exhibit a classic three-component structure \citep{1985JGR....90..275I, 2012ApJ...752...36C, 2012ApJ...750...12S, 2023ApJ...942...19S} that appears in the outer corona as a bright front, followed by a darker cavity that frequently contains a bright core, with typical angular widths around 50\degr -- 60\degr \citep{2014GeoRL..41.2673G,2021FrASS...8...73P}. However, not all CMEs have the typical three-component structure. In more cases, they appear as narrow-band (or jet-like) \citep[e.g.,][]{2005ApJ...628.1056L,2008SoPh..249...75L,2022SoPh..297..138Z,2023ApJ...952...85W}, or structureless bubble-like\citep[e.g.,][]{2013SoPh..284..179V,2017ApJ...838..141V} or even as CMEs with both forms (double-structure CMEs, abbreviated as twin CMEs)\citep[e.g.,][]{2012ApJ...745..164S,2018ApJ...869...39M,2019SoPh..294...68S,2019ApJ...881..132D}. This seems to indicate that the mechanism underlying CME production is still unclear. So why are there so many different forms of CMEs in the high corona, and what determines the characteristic features of CMEs in the high corona? Among all of these CMEs, twin CMEs are very unique phenomena that refer to the simultaneous production of jet-like and bubble-like components in one single CME event. This kind of CME event was first reported by \citet{2012ApJ...745..164S} and subsequently by \citet{2018ApJ...869...39M,2019SoPh..294...68S,2019ApJ...881..132D}. Studying twin CMEs can help us understand the mechanism behind the production of these different types of CMEs.

\par
Coronal jets are another type of plasma ejection activity phenomenon in the solar corona \citep{2021RSPSA.47700217S}. Early observations indicated that the bright points at the base of coronal jets seemed to provide important insights into their energy source \citep{1992PASJ...44L.173S}.  \citet{2010ApJ...720..757M} referred to coronal jets with a shear core field at their base as “blowout jets". Recently, more and more observational studies reported that the blowout jets were often associated with eruptive small filaments \citep[e.g.,][]{2011ApJ...738L..20H, 2012ApJ...745..164S, 2015Natur.523..437S, 2015ApJ...815...71C, 2017ApJ...851...67S, 2018ApJ...854...80H, 2019ApJ...885L..11S, 2020ApJ...900..158Y, 2023ApJ...945...96Y, 2023ApJ...953..171H}. Considering the three-dimensional (3D) scenario, with the aid of magnetic field extrapolation techniques and high-resolution observations, an increasing number of studies revealed that the 3D magnetic configuration of the coronal jets is a fan-spine structure\citep{2021RSPSA.47700217S}. This structure is composed of a null point, a dome-shaped fan, and the inner (outer) spines located inside (outside) the fan, which is typically accompanied by a circular flare ribbon and an inner bright patch \citep[e.g.,][]{2009ApJ...700..559M, 2012ApJ...760..101W, 2017ApJ...842L..20L, 2018ApJ...859..122L, 2019ApJ...885L..11S, 2019ApJ...871....4H, 2020ApJ...893..158L, 2022ApJ...926L..39D, 2024ApJ...962L..38D, 2024RvMPP...8....7Z,2024ApJ...965..137H}. In a high-resolution MHD simulation, \citet{2017Natur.544..452W, 2018ApJ...852...98W} showed a breakout model \citep{1999ApJ...510..485A} for coronal jets in a 3D fan-spine structure involving small filaments. The instability and eruptions of the filament may be caused by the photospheric flux emergence and/or cancellation \citep[e.g.,][]{2012ApJ...745..164S, 2014ApJ...783...11A, 2017ApJ...844..131P, 2018ApJ...869...78C, 2020ApJ...902....8C, 2021ApJ...912L..15T}, or breakout reconnection \citep[e.g.,][]{2017Natur.544..452W, 2018ApJ...852...98W, 2018ApJ...854..155K, 2019ApJ...873...93K, 2023ApJ...953..148S, 2024MNRAS.528.1094Y}. The jets are produced by the latter process as the filament flux reconnects with and releases twists into the external magnetic flux. The relationship between coronal jets and Type \rmnum{3} radio bursts has also been extensively discussed \citep[e.g.,][]{1995ApJ...447L.135K, 1996A&A...306..299R, 2011A&A...531L..13I, 2011SoPh..273..413K, 2013ApJ...763L..21C, 2017ApJ...835...35H, 2018ApJ...861..105S, 2021A&A...647A.113Z, 2022ApJ...926L..39D, 2022ApJS..260...19Z}. Type \rmnum{3} radio bursts are a crucial diagnostic tool for understanding the acceleration of electrons associated with outward-propagating coronal jets \citep{2020FrASS...7...56R}. The leading theory suggests that, since magnetic reconnection occurs between closed and open magnetic fields, electrons are accelerated by reconnection and subsequently escape along open magnetic field lines into interplanetary space, emitting Type \rmnum{3} radio bursts.

\par
Considering the relationship between coronal jets and CMEs, \citet{1998ApJ...508..899W} discovered that the narrow-band CMEs were the extensions of EUV jets in the high corona. The disturbances in the closed magnetic fields contained the twist that led to its reconnection with the surrounding open field, resulting in the release of energy in the form of a jet  \citep[e.g.,][]{2009SoPh..259...87N, 2011ApJ...738L..20H, 2016ApJ...819L..18Z}. Jets extended to interplanetary space, which appear as enhanced density in the white-light coronagraph images as the narrow CMEs. \citet{2016ApJ...822L..23P} suggested that the small filament twisted at the base of the jet transfer its twist to large-scale magnetic loops and drives these loops outward \citep[e.g.,][]{2015ApJ...813..115L} along a streamer, forming “Streamer-pull CMEs". Furthermore, \citet{2021ApJ...911...33C} reported that rapid magnetic twisting at the base of the jet can even lead to the formation of large-scale magnetic flux ropes, which erupt outward and result in bubble-like CMEs towards Earth's space. Considering the three-dimensional fan-spine magnetic field structure, \citet{2021ApJ...907...41K} contended that whether a jet or a CME is produced is determined by the ratio between the magnetic free energy of the filament channel and the magnetic free energy inside and outside the dome-shaped fan of the fan-spine configuration. Interestingly, \citet{2021ApJ...909...54W} simulated a complex magnetic field configuration in which the twist confined in a pseudostreamer topology is not only transferring twist into open field lines as a jet but also partially injected into the closed field beneath the adjacent helmet streamer. This injected twist blew off the closed streamer top to result in a complex CME, which consisted of a mixture of open and closed magnetic field lines.

\par
To better clarify the physical factors determining the morphology of jet-related CMEs, we conducted an analysis of the source region for seven twin CME events (three of which were previously studied) and nine narrow CME events from October 2010 to December 2012. Using magnetic field extrapolation and/or three-dimensional reconstruction, we obtained the initial null point height. Additionally, we measured various parameters, including filament length, CME velocity, and other relevant factors. It should be noted that to avoid confusion, hereafter, we employ the term “jet-like CME" to indicate one component of a twin CME, while the term “narrow CME" refers to a CME caused by one jet. Moreover, we note that pseudostreamers (as simulated by \citet{2021ApJ...909...54W}) are similar to but more complex than fan-spine structures, with multiple nulls, separators, and open outer spines. In this paper, we describe the structure as a fan-spine configuration, noting that due to observational constraints, we can only determine the presence of at least one magnetic null point, without excluding the possibility of multiple magnetic null points in actual scenarios.

\section{Observations And Data Analysis} \label{sec:instr}
We combined the data from Solar Dynamics Observatory (SDO; \citet{2012SoPh..275....3P}) with that from Solar Terrestrial Relations Observatory (STEREO, \citet{2008SSRv..136....5K}), focusing on jet-related twin CMEs occurring from 2010 October 1 to 2012 December 31.  The separation angle between the two STEREO spacecraft and SDO is between 80$^{\circ}$$-$150$^{\circ}$. As the viewing angle at this period is optimal, it is easy to rule out the possibility that the twin CMEs are not caused by a single jet. Otherwise, there are many twin CMEs in the Large Angle and Spectrometric Coronagraph (LASCO; \cite{1995SoPh..162..357B}), but they are impacted by projection or caused by multi-jets, rather than the twin CMEs described in our paper. The common requirement that the jets be outside of 30$^\circ$ east-west longitude and beyond 20$^\circ$ north-south latitude was the basis for our selection of the events. Then, we checked all of the jets in SDO/AIA 304 \AA\ using this criterion. Finally, we confirmed the morphology of the CMEs in LASCO/C2 and verified whether their occurrence time and corresponding locations were consistent with the jets. 
\par
If the CMEs exhibited two components (jet-like and bubble-like), we need to confirm in STA/STB whether the twin CME was caused by a single jet. We introduced four best twin CME cases and three events mentioned in the previous literature \citep[e.g.,][]{2012ApJ...745..164S,2019SoPh..294...68S,2019ApJ...881..132D}, merging them with nine narrow CMEs for comparison.
Be aware that \cite{2018ApJ...869...39M} also recorded a twin CME event. However, due to the source jet being positioned near the edge of the active region and obscured by loops, no analysis of this event is included in the list presented herein. 
\par
For this study, we mainly used 304 \AA~images from the Atmospheric Imaging Assembly (AIA; \citet{2012SoPh..275...17L}) on board the SDO to study the jets. The images from the C2 coronograph of the LASCO onboard the Solar and Heliospheric Observatory (SOHO) were used to study the CMEs. Combined with the Extreme Ultraviolet Imager (EUVI; \citet{2004SPIE.5171..111W}) and COR1 on board the two STEREO spacecraft , we also applied these observations to rule out the possibility that the twin CME did not come from one single jet.
\par
The SDO/AIA (STEREO/EUVI) 304 \AA~images utilized in this study possess a pixel resolution of 0\arcsec.6 (1\arcsec.56) and a cadence of 12s (10 min). For the LASCO/C2, the field of view (FOV) is from 2 to 6 solar radii, and the images are captured at a cadence of 12 minutes. The STEREO/COR1 has a temporal cadence of 5 minutes and provides observations of the outer corona within the range of 1.4 to 4 solar radii. In addition, Type \rmnum{3} bursts are detected by the WAVES instrument \citep{1995SSRv...71..231B} on board the WIND spacecraft, operating within a frequency range of 0.02$-$13.825 MHz.
\par
Taking advantage of the Potential Field Source Surface (PFSS, \citet{2003SoPh..212..165S}) magnetic extrapolation technology provided by the SSW, we checked the magnetic field configuration before (or after) the eruption for all events. As a supplement, we also chose the line-of-sight HMI magnetogram of the event source region as the bottom boundary and utilized the code provided by SolarSoftware (SSW, \cite{1998SoPh..182..497F}) for local potential field extrapolation. Of course, these early (or later) extrapolations can’t account for the emergence or other changes that could occur between those times and the onset of the limb events, but the extrapolations would yield more solid evidence of the fan-spine configurations as well as a better understanding of the surrounding fields \citep[e.g.,][]{2021ApJ...907...41K}.

\section{RESULTS} \label{sec:OBS}
\subsection{Overview of Twin CMEs}
The four twin CME events that we are interested in, along with their corresponding source jets, are displayed in Figure 1. The feature of filaments located at the base of Jet1$-$Jet4 (J1$-$J4) in the top two rows is highlighted by crosses. The lengths of the filaments were obtained using a three-dimensional (3D) reconstruction method, as detailed in Section 3.7. As we can see, J1, J2, and J4 are close to the solar limb in Figure 1 (a1)$-$(a4) images, while J3 is on the solar disk. Combined with the different perspectives offered by the STEREO/EUVI (see Figure 1 (b1)$-$(b4)), it provides a unique opportunity to explore the underlying factors that govern the structural characteristics of jet-related twin CMEs. The twin CMEs present two components, namely a jet-like CME and a bubble-like CME, which is consistent with the characteristic morphology often associated with twin CMEs \cite[e.g.,][]{2012ApJ...745..164S}. A detailed analysis of the eruption source region was undertaken to explore the possible mechanism contributing to the formation of twin CMEs. This investigation focused on four events, as presented in sections 3.2, 3.3, 3.4, and 3.5. The analysis was carried out using two perspectives, SDO and STEREO, to provide a comprehensive understanding.  

\begin{figure}
\centerline{\includegraphics[width=0.45\textwidth,clip=]{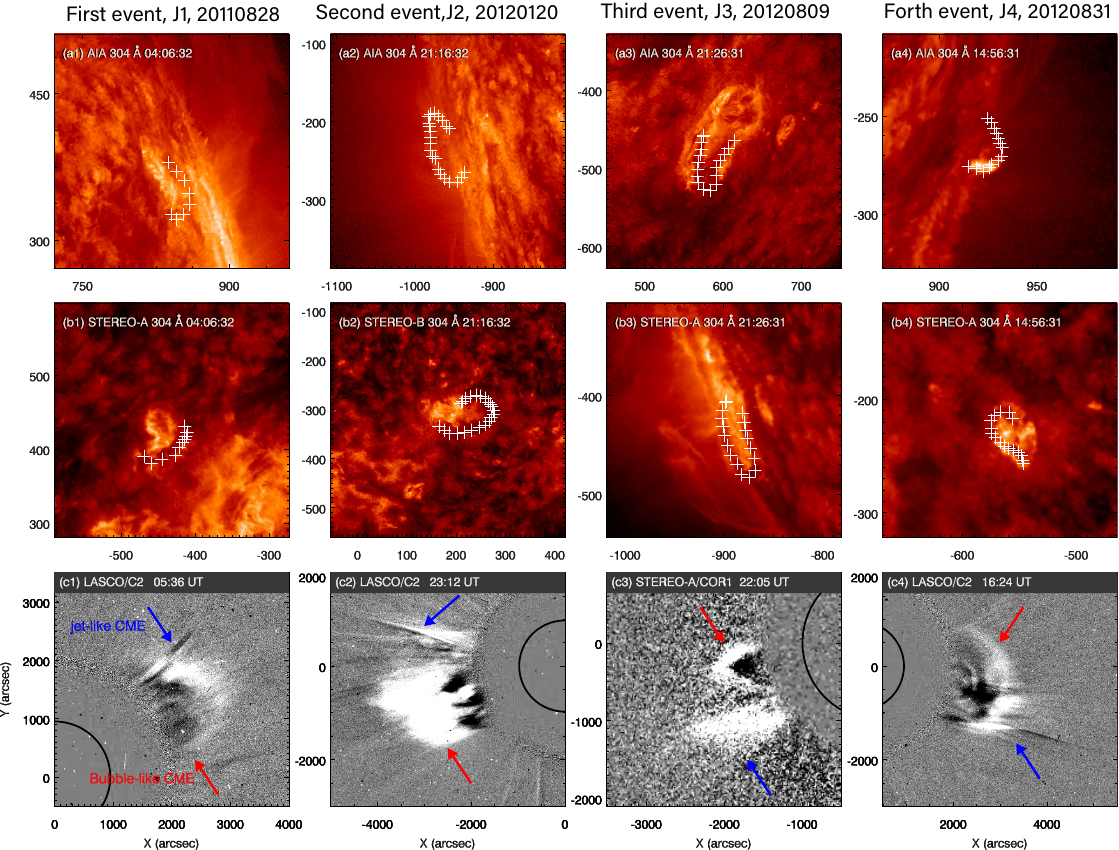}}
\caption{The overview of four twin CME events. (a1)$-$(a4): AIA 304 \AA~images show the source location of four solar jets, J1, J2, J3, and J4. (b1)$-$(b4) provide the corresponding source region in STEREO/EUVI 304 \AA~. The main structure of the filaments is indicated with crosses in all images. (c1)$-$(c4): the coronagraph running difference images displaying the twin coronal mass ejections (CMEs). The jet-like CME and bubble-like CME are denoted by blue and red arrows, respectively.} 
\label{fig1}
\end{figure}

\subsection{2011 August 28}
\par Figures 2, 3, and 4 present detailed observations of the first event, J1, which occurred on 2011 August 28. Figure 2 illustrates the dynamic formation process of a jet from the perspective of STEREO-A. As seen in Figure 2 (b1), there is a circular ribbon present before a filament going up. When the filament lifts at 04:25 UT, an inner bright patch can be identified in Figure 2 (b2). These characteristics suggest that the filament is situated within a fan-spine magnetic field configuration, as suggested by previous works \citep[e.g.,][]{2019ApJ...885L..11S,2019ApJ...872...87L,2020ApJ...898..101Y,2021ApJ...923...45Z,2022ApJ...926L..39D,2024ApJ...962L..38D}. The filament flux reconnects with and releases twist into the external flux, producing a jet, as shown in Figure 2 (a3)$-$(c3). This magnetic configuration of the filament changes from closed to open flux, indicating the occurrence of the magnetic reconnection between the filament flux and the ambient opposite-polarity magnetic flux. Interestingly, during this reconnection process, an upward-moving EUV wave is observed ahead of the jet, as the cyan arrows and dashed line shown in Figure 2 (c2)$-$(c3). Then, the material falls back to the solar surface along the surrounding magnetic field lines (see cyan arrows in Figure 2 (a4) and (c4)). 

\begin{figure}
\centerline{\includegraphics[width=0.45\textwidth, clip=]{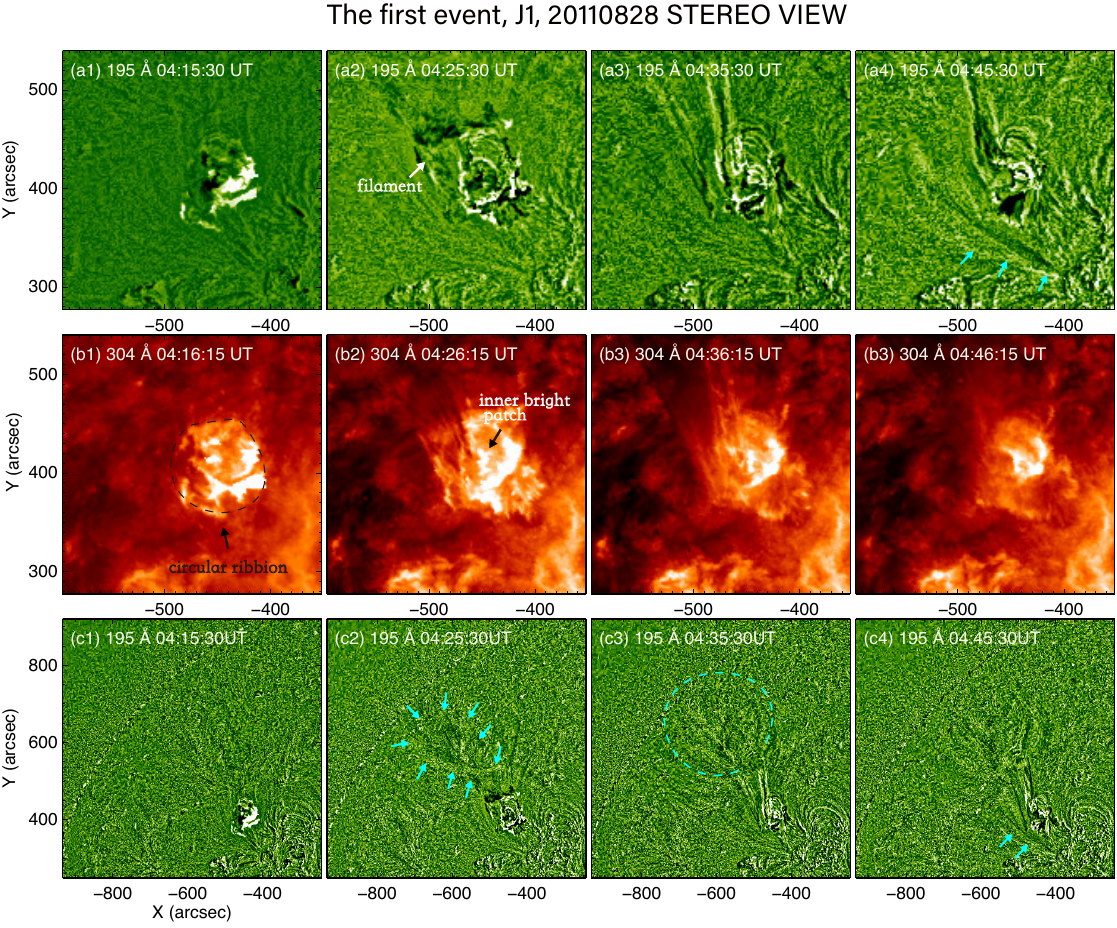}}
\caption{The first event, J1, is from 2011 August 28 on STEREO-A view. (a1)$-$(a4) and (c1)$-$(c4) the running difference images of EUVI 195 \AA~. The cyan arrows and dashed line indicate an upward-moving wave ahead of a jet. In (a4) and (c4), the cyan arrows point to the location where the material falls back to the solar surface along the magnetic field lines. An animation of the running difference images of EUVI 195 \AA~but at a larger field-of-view, is available from 03:05 UT to 06:50 UT. The duration of this animation is 1 s.}
\label{fig2}
\end{figure}

\begin{figure}
\centerline{\includegraphics[width=0.45\textwidth,clip=]{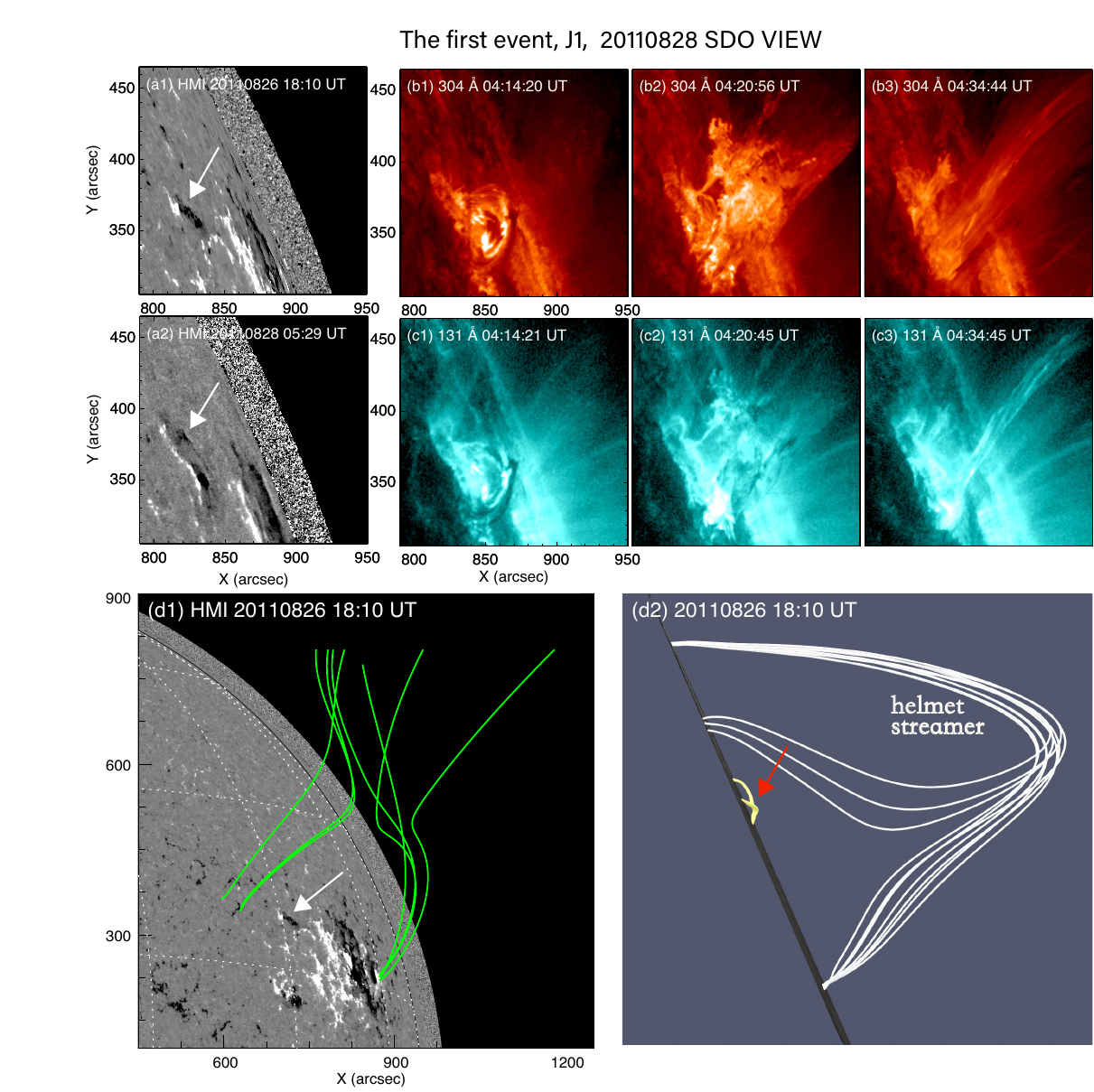}}
\caption{Photospheric magnetograms and EUV images from SDO, and potential field extrapolations, for the source region of J1. (a1)$-$(a2) the images of the magnetic field before and after the jet eruption, respectively. (b1)$-$(b3) and (c1)$-$(c3): the sequence images taken by AIA 304 \AA~and 131 \AA~. In (d1)$-$ (d2), the extrapolated field lines overlying the HMI magnetogram in 2011 August 26, two days before the jet eruption. The white arrows denote the eruptive region in panels (a1), (a2) and (d1). The green lines represent the open field lines. The red arrow points out he null point location and the adjacent helmet streamer is represented by the white lines. An animation of the evolution of the magnetic flux (a1) is available from 2011 Aug 26 09:34 UT to 2011 Aug 28 04:46 UT. The duration of this animation is 9 s.}  
\label{fig3}
\end{figure}

\par Furthermore, the perspective provided by the SDO offers us further insights for the initial launch of the jet. Figure 3 (a1)$-$(a2) shows the variations in the photospheric magnetic field for the event J1 from the SDO/AIA view. The filament within the fan-spine topology rises for some reason (potentially resulting from magnetic flux cancellation \citep[e.g.,][]{2016ApJ...822L..23P}, shear motion \citep[e.g.,][]{1999ApJ...510..485A,2020ApJ...902....8C}, or a kink-like instability of the stressed internal closed field \citep[e.g.,][]{2009ApJ...691...61P,2017ApJ...834...62K} and subsequently erupts, forming a jet along the magnetic field lines, as shown in the 304 and 131 \AA~sequence images in Figure 3. It should be noted that the magnetic flux of this example emerged near the solar limb approximately 2 days earlier (see the animation accompanying Figure 3). We also utilized the PFSS extrapolation technique to check the magnetic field configuration before the J1 eruption, as shown in Figure 3 (d1). The local potential-field extrapolation shows the emerging region is adjacent to a helmet streamer \citep[e.g.,][]{2007ApJ...660..882W,2023ApJ...952...85W}, as illustrated in Figure 3 (d2). This topological configuration is consistent with the simulations conducted by \citet{2021ApJ...909...54W}.

\begin{figure} 
\centerline{\includegraphics[width=0.45\textwidth,clip=]{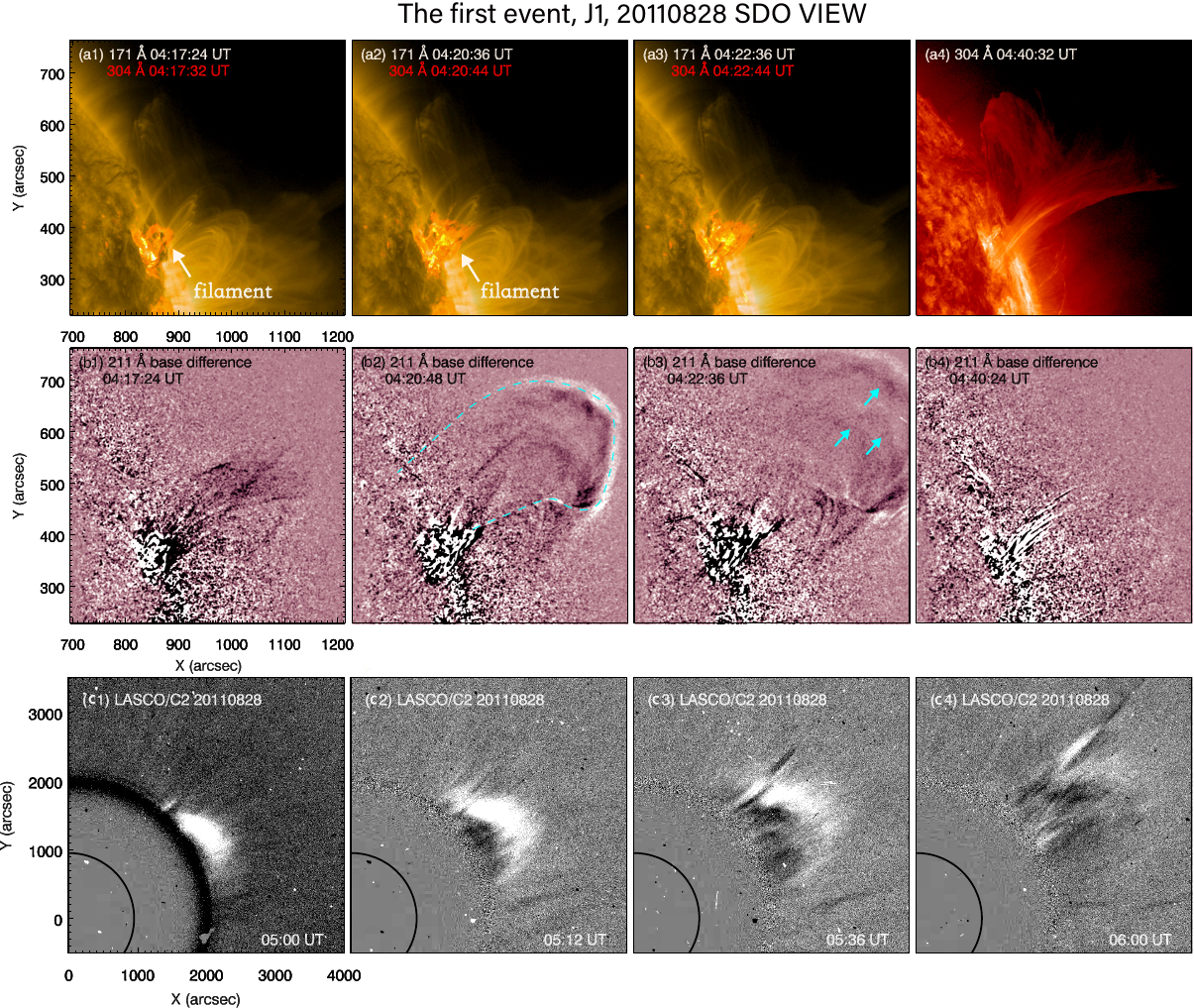}}
\caption{The dynamic process of jet-related twin CME formation for event J1. (a1)$-$(a3) the superimposed images of AIA 171 \AA~ and 304 \AA~. (b1)$-$(b4) the difference sequence images of AIA 211 \AA~. The cyan arrows point to an upward-moving wave ahead of the jet. The dashed line outlines the upward-moving wave, agreeing with Figure 2 (c3). (c1)$-$(c4) the evolution of twin CME from LASCO/C2. An animation of the diifference sequence images of AIA 211 \AA~is available from 04:00 UT to 04:49 UT. The duration of this animation is 4 s.}
\label{fig4}
\end{figure}

\par Before the eruption of the filament, multiple sets of magnetic loops were observed near the filament in AIA 171 \AA~, as shown in Figure 4 (a1). Interestingly, at 04:20 UT, as the filament starts to rise and forms a jet, an upward-moving wave can be seen heading ahead of the jet, as indicated by the dashed line in Figure 4 (b2) and the cyan arrows in Figure 4 (b3). Combining the observational results from the STEREO-A view in Figure 2 (c2)$-$(c3) and the magnetic field configuration depicted in Figure 3 (d1)$-$(d2), it is anticipated that this upward-moving wave is related to the helmet streamer located adjacent to the fan-spine jet, as simulated by \citet{2021ApJ...909...54W}. It should be noted that the fan-spine jet and this helmet streamer system originate in different parts of the PIL and the different parts of the PIL have different amounts of stored free energy and/or overlying flux. In other words, besides transferring onto open field lines and forming the jet, a portion of the twist is injected into the closed field beneath the neighboring helmet streamer. The twist added to the open field then propels the jet to propagate into interplanetary space and manifest as a jet-like CME observed in LASCO/C2 coronagraph images. The injected twist drives the adjacent helmet streamer outward, forming the bubble-like CME, as shown in Figure 4 (c1)$-$(c4). These observational results support the scenario simulated by \citet{2021ApJ...909...54W} and the recent observation from \citet{2021ApJ...911...33C}.

\subsection{2012 January 20} Figure 5 displays the second event, J2, occurring at the solar disk on 2012 January 20 from STEREO-B. One can clearly identify a circular ribbon that manifested the footpoint locations of the fan structure. This observation suggests that the jet was driven by the filament eruption inside a fan-spine configuration.

\par The process of jet formation can be observed more clearly from the perspective of SDO/AIA, as displayed in Figure 6 (a1)$-$(a3) and (b1)$-$(b3). A filament undergoes reconnection with surrounding open magnetic fields, resulting in the outward ejection of a jet. The outward trajectory of the jet, away from the solar surface, manifests as a jet-like component of twin CME in the higher solar corona (see Figure 6 (c2) and (c3)). It is intriguing that during the eruption process of the filament, an eastward-moving wave becomes discernible, as shown in Figure 6 (b2). To determine the magnetic field configuration of the J2 eruption, we conducted a magnetic field extrapolation of the eruption source region four days later, as illustrated in Figure 6 (d1)$-$(d2). Figure 6 (d1) displays open field lines surrounding the source region, and the local potential field extrapolation in (d2) shows the presence of a helmet streamer surrounding the fan dome. Considering the filament within the fan-spine system, it can be speculated that the formation of the bubble-like component in this event (see Figure 6 (c1)$-$(c3)) exhibits similarities to the description provided in Section 3.2.

\begin{figure}[b]
\centerline{\includegraphics[width=0.45\textwidth,clip=]{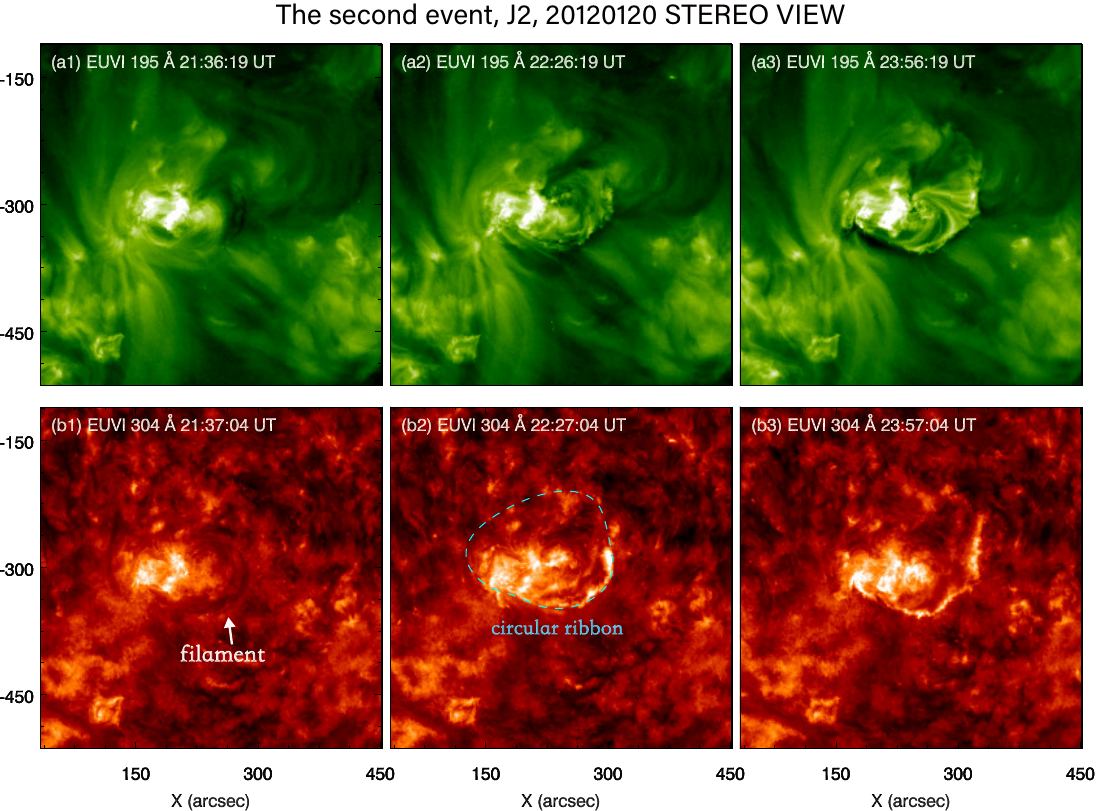}}
\caption{The second event, J2, is from 2012 January 20 on STEREO-A view. (a1)$-$(a3) and (b1)$-$(b3) the sequence images taken by EUVI 195 \AA~and 304 \AA~, respectively. In (b2), the cyan dashed line outlines the circular ribbon before the filament eruption.}
\label{fig5}
\end{figure}

\begin{figure}
\centerline{\includegraphics[width=0.45\textwidth,clip=]{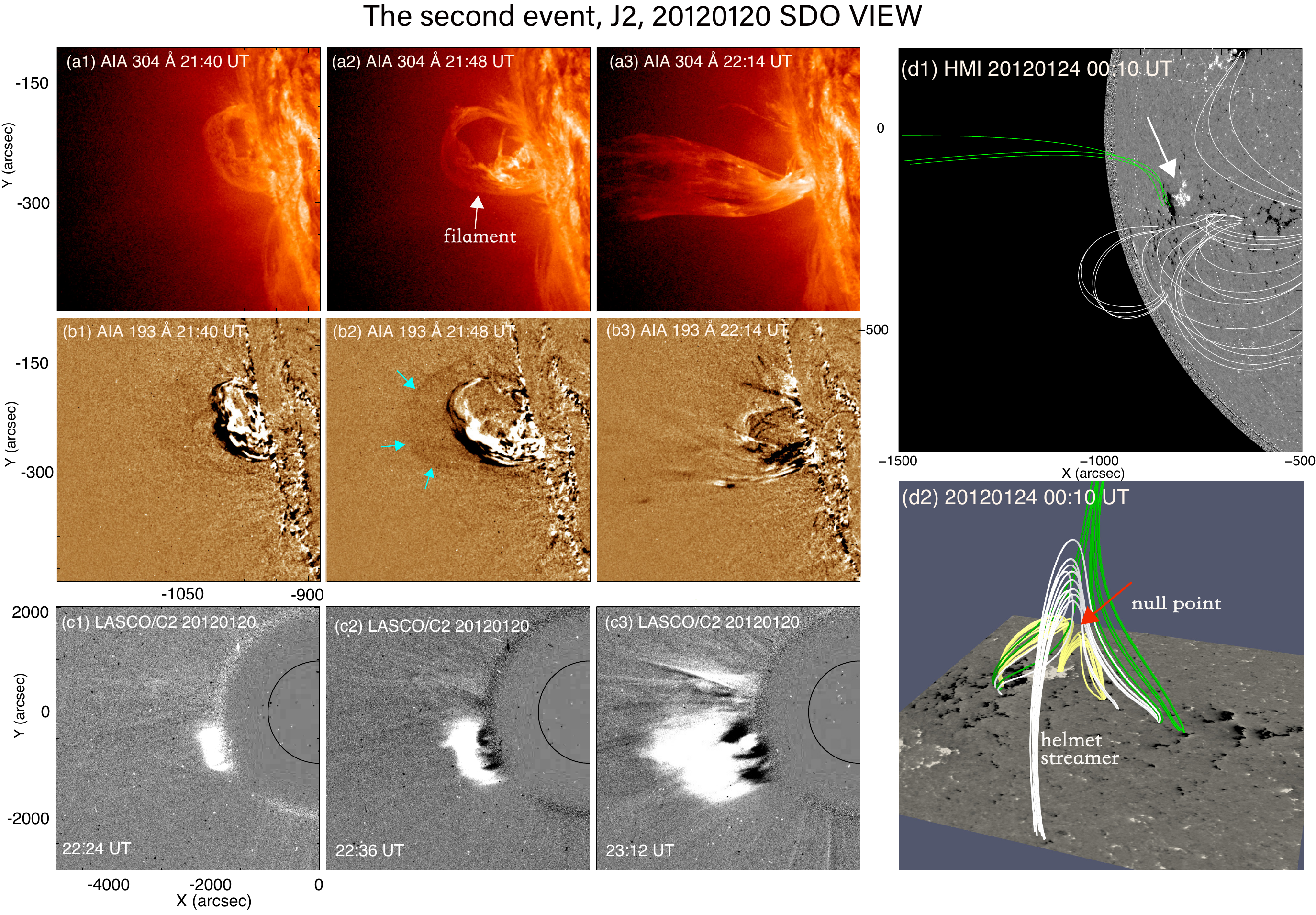}}
\caption{Photospheric magnetograms and EUV images from SDO, the twin CME from LASCO, and potential field extrapolations for the source region of J2. (a1)$-$(a3) and (b1)$-$(b3): a sequence image of AIA 304 \AA~ and 193 \AA~ running difference, respectively. (c1)$-$(c3) the evolution of twin CME from LASCO/C2. In (d1)$-$(d2), the extrapolated field lines overlaid on the HMI magnetograms on 2012 January 24, four days after the jet eruption. The white arrows denote the eruptive region. An animation of the AIA 193 \AA sequence is available from 21:31 UT to 22:29 UT. The duration of this animation is 5 seconds.}
\label{fig6}
\end{figure}

\subsection{2012 August 9}
\par Figure 7 presents the best example within our event, termed J3, wherein the magnetic fields in the eruption source region can be identified from the HMI magnetogram on 2012 August 9 at 18:00 UT (see Figure 7 (a1)). Prior to the eruption, the three-dimensional (3D) coronal magnetic field structure of J3 was obtained through extrapolation using the PFSS model. The relevant extrapolated magnetic field lines were superimposed on the HMI magnetogram, revealing a fan dome encompassed by both open and closed (helmet streamer) flux, as shown in Figure 7 (a1). Interestingly, the potential-field extrapolation exhibits a fan-spine topology, comprised of an inner spine and open outer spines, as represented by yellow and green lines in Figure 7 (d1) and (d2)). The fan-spine system resides in the neighboring helmet streamer, with the closed field lines depicted by white lines in Figure 7 (d1) and (d2). As simulated by \citet{2021ApJ...909...54W}, the dome connects the fan-spine topology to the nearby helmet streamer, so that reconnection between the fan-spine system and helmet streamer fluxes is possible. Figure 7 (b1)$-$(b3) and (c1)$-$(c3) are a close-up view of the filament eruption in different wavelengths, accompanied by a big field of view of a jet (see Figure 7 (a2) and (a3)). Initial, obvious brightening arose right below the north footpoint of the filament (see Figure 7 (b1) and (c1)). Afterward, the filament lifted displaying a twisted shape, and reconnected with the surrounding open magnetic field lines. At 21:50 UT, it can be observed that the material of a jet is ejected outward along these open magnetic fields, as pointed to by yellow arrows in Figure 7 (a2) and (b2). It can be observed that the initial brightening is at the north end of the PIL (Figure 7 (b1) and (c1)), while the post-flare loops appear at the south end (denoted by the pink arrow in Figure 7 (c2) and (c3)), indicating that the filament lifts first at the north end but erupt more fully at the south end. And a cusp structure becomes clearly visible after the filament eruption. These apparent topological changes provide evidence of the reconnection process occurring between the erupting twisted filament and the surrounding opposite fields. 

\begin{figure}
\centerline{\includegraphics[width=0.45\textwidth,clip=]{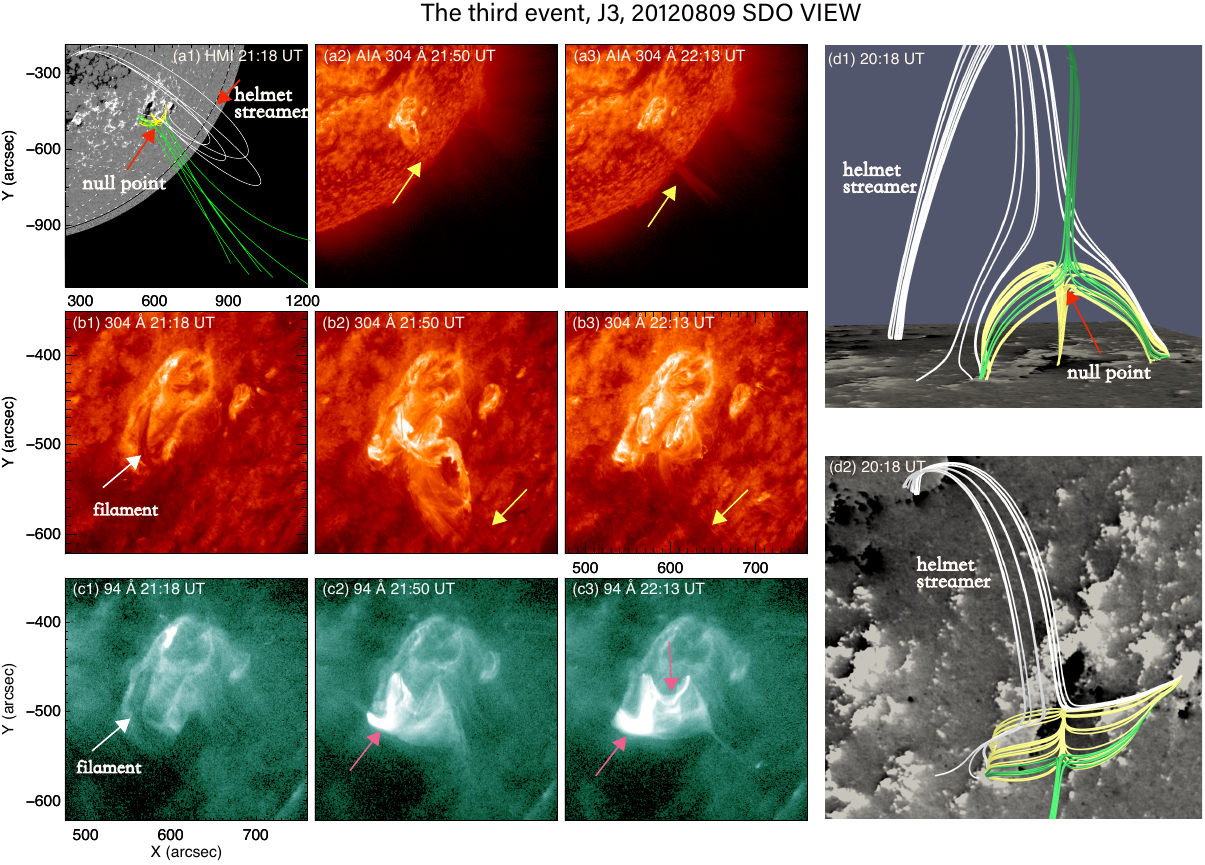}}
\caption{The third event, J3, is from 2012 August 9. (a1) the extrapolated field lines obtained from the PFSS model are overlaid on the HMI magnetogram. (a2)$-$(a3) the sequence images of AIA 304 \AA~. A close-up view of the evolution of case J3 is displayed in (b1)$-$(b3) and (c1)$-$(c3). The yellow arrows point to the jet along open fields, and the pink arrows denote the post-flare loops. The yellow (green) lines represent the closed (fan) and open field lines in panels (a1), (d1) and (d2). The red arrows point out the null point and the white lines represent the adjacent helmet streamer. An animation of the jet eruption, panels (a2) and (a3), is available from 21:09 UT to 22:49 UT. The duration of this animation is 8 s. }
\label{fig7}
\end{figure}

\par The STEREO/EUVI from another perspective reveals that the helmet streamer is situated near the filament, as shown in Figure 8 (b1)$-$(b3)). Notably, a jet-like CME appeared in the COR1 field of view at a distance of 1.4-4.0 solar radii (as seen in Figure 8 (c1)) shortly after the filament began to rise at 21:40 UT (as shown in Figure 8 (a1)). This suggests that the material comprising the initial jet-like CME may have originated from the magnetic reconnection between the confining field above the filament and the surrounding open magnetic fields, which is consistent with \citep{2012ApJ...745..164S}. As time progresses, the filament located within the fan-spine topology gradually rises and undergoes reconnection with the surrounding magnetic fields (the details are shown in Figure 7). Consequently, an ejected jet propagates along the open field lines of the outer spine of the fan-spine system, as shown in Figure 8 (a3), which is consistent with the mentioned result of SDO/AIA. These results demonstrate that the twist contained in the fan-spine system passes into the open fields, allowing it to manifest as a jet-like CME in STEREO/COR1 coronagraph images.
\par It is worth noting that when a bubble-like CME first became discernible in the COR1 field of view at 21:50 UT in Figure 8 (c2), the filament was observed to be in the ascending stage, which is shown in Figure 7 (a2)$-$(b2) and Figure 8 (a2). This result suggests that the bubble-like CME does not directly result from the filament eruption, which is different from the interpretation of \citep{2012ApJ...745..164S}. As the filament lifts, magnetic reconnection occurs at the null point (also known as breakout reconnection), and it can be observed that material falls back to the solar surface along closed magnetic field lines, as indicated by the cyan arrows in Figure 8 (b2) and (b3). This indicates that the fan loops undergoing reconnection are part of the helmet streamer. Combined with the potential-field extrapolation in Figure 7, these results strongly support the scenario that was proposed in the simulations conducted by \citet{2021ApJ...909...54W}. The twist contained within the fan-spine topology does not transfer entirely onto open field lines as in a jet. Apart from that, a significant portion of the twist is injected into the closed field beneath the adjacent helmet streamer. Thus, this result leads to the loops being blown outward and producing a bubble-like CME in the high corona. In this case, we did not observe the bubble source in the EUVI image, possibly because it is behind the limb from that point of view.

\begin{figure}
\centerline{\includegraphics[width=0.45\textwidth,clip=]{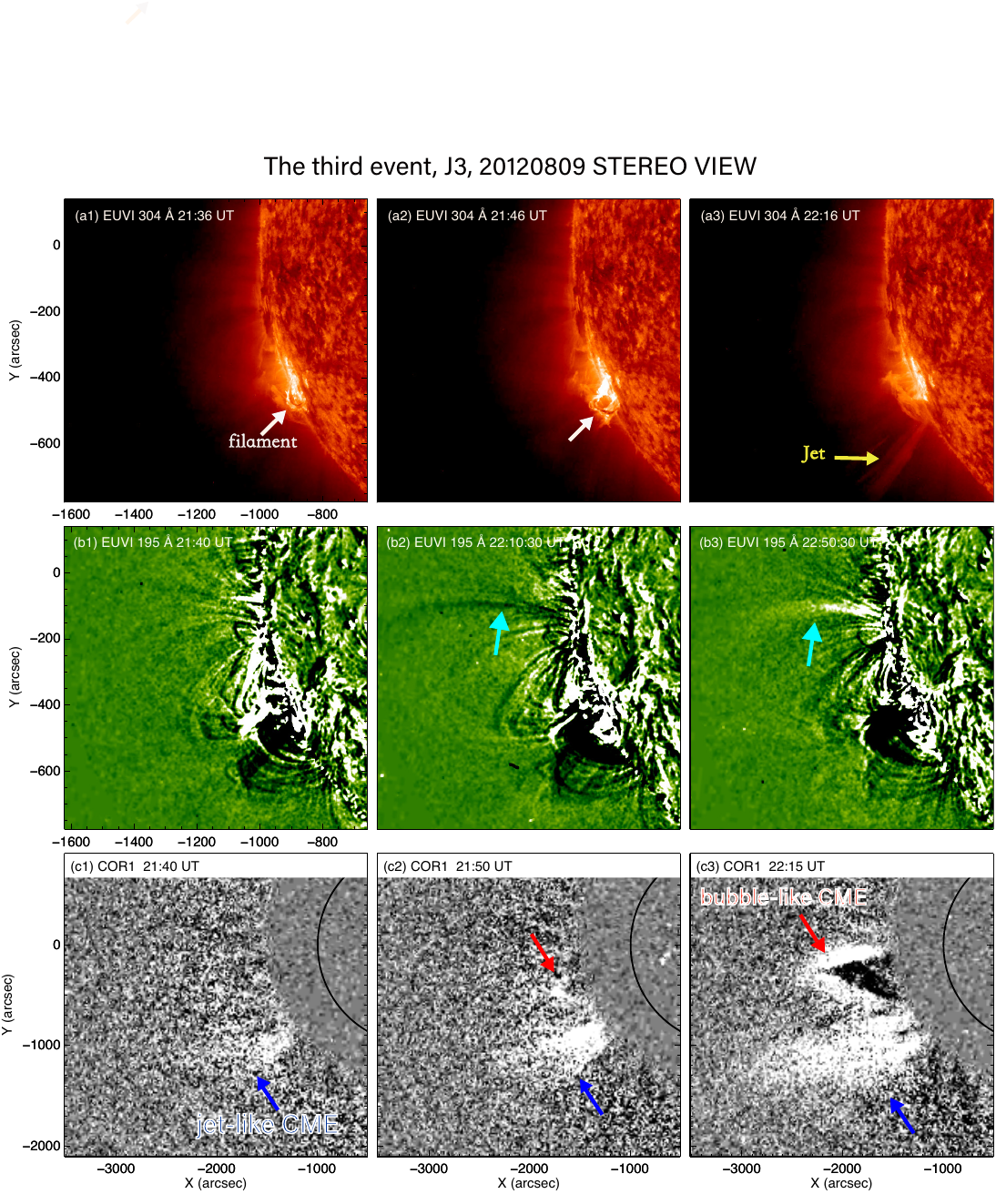}}
\caption{The dynamic process of jet-related twin CME formation from STEREO-A view on 2012 August 9. The cyan arrows point to the position where the material returns to the solar surface along the magnetic field lines in (b2)$-$(b3). The twin CME is from STEREO/COR1 in (c1)$-$(c3). An animation of the twin CME, panels (c1) - (c3), is available from 21:05 UT to 23:55 UT. The duration of this animation is 2 s.}
\label{fig8}
\end{figure}

\begin{figure}
\centerline{\includegraphics[width=0.45\textwidth,clip=]{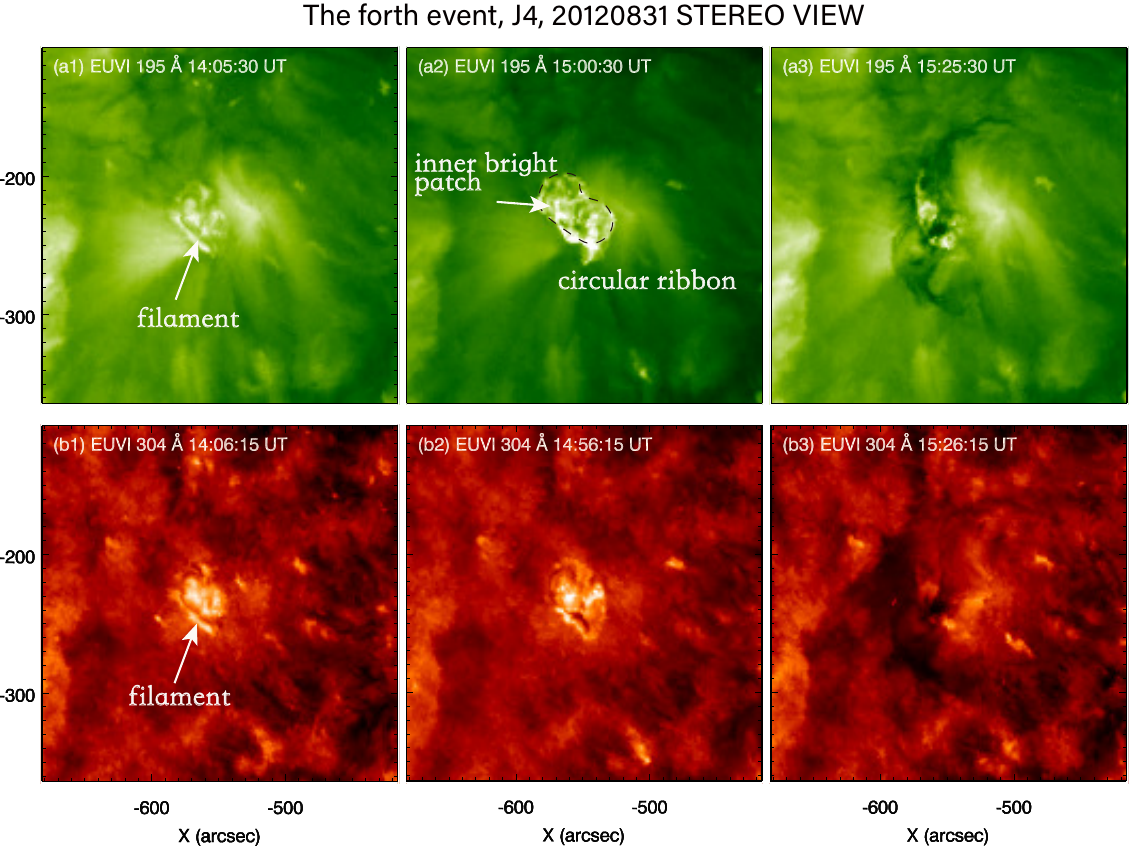}}
\caption{The fourth event, J4, is from 2012 August 31 on STEREO-A view. (a1)$-$(a3) and (b1)$-$(c3) the images of EUVI 195 \AA~and 304 \AA~,respectively. The black dashed line outlines the circular ribbon before the filament eruption. }
\label{fig9}
\end{figure}

\subsection{2012 August 31}
Figures 9 and 10 display the results of event J4 from 2012 August 31. Initially, Figure 9 (a1)$-$(b1) displays a hook-shaped filament. Afterword, before the filament eruption, a circular ribbon can be observed in Figure 9 (a2), outlined by a black dashed line. This agrees well with the eruptive characteristic observed under the three-dimensional fan-spine magnetic configuration. 
\par
By using a large filament surrounding J4 as a tracer, we discovered that the magnetic configuration of J4 resembled a fan-spine structure one day before the eruption, as shown in Figure 10 (a1). HMI magnetogram shows that J4 originates from a location where negative polarity magnetic flux emerges into positive polarity magnetic flux. Based on local potential field extrapolation, it can be observed that one day before, the magnetic field structure of J4 exhibits a fan-spine topology, as displayed in Figure 10 (d1) and (d2). This suggests that the J4 observed one day later occurs under the same fan-spine system. Interestingly, it can be observed that the fan dome of this fan-spine structure is located on one side of a large helmet streamer, but the spine of this system is outside the helmet streamer, extending into interplanetary space. During the eruption of J4, it can be observed that a set of loop systems surrounds it, as pointed out by cyan arrows in Figure 10 (b2). It is highly likely that the set of loop systems corresponds to the helmet streamer that was demonstrated in the potential field extrapolation one day before. The formation of a bubble-like component of twin CME in Figure 10 LASCO/C2 images shares similarities with the descriptions mentioned in Sections 3.2, 3.3, and 3.4, indicating a portion of twist is injected into the closed helmet streamer that blows out the top of the streamer, as simulated by \citet{2021ApJ...909...54W}. The jet-like component represents a natural extension of the jet along the open magnetic field line.

\begin{figure}
\centerline{\includegraphics[width=0.45\textwidth,clip=]{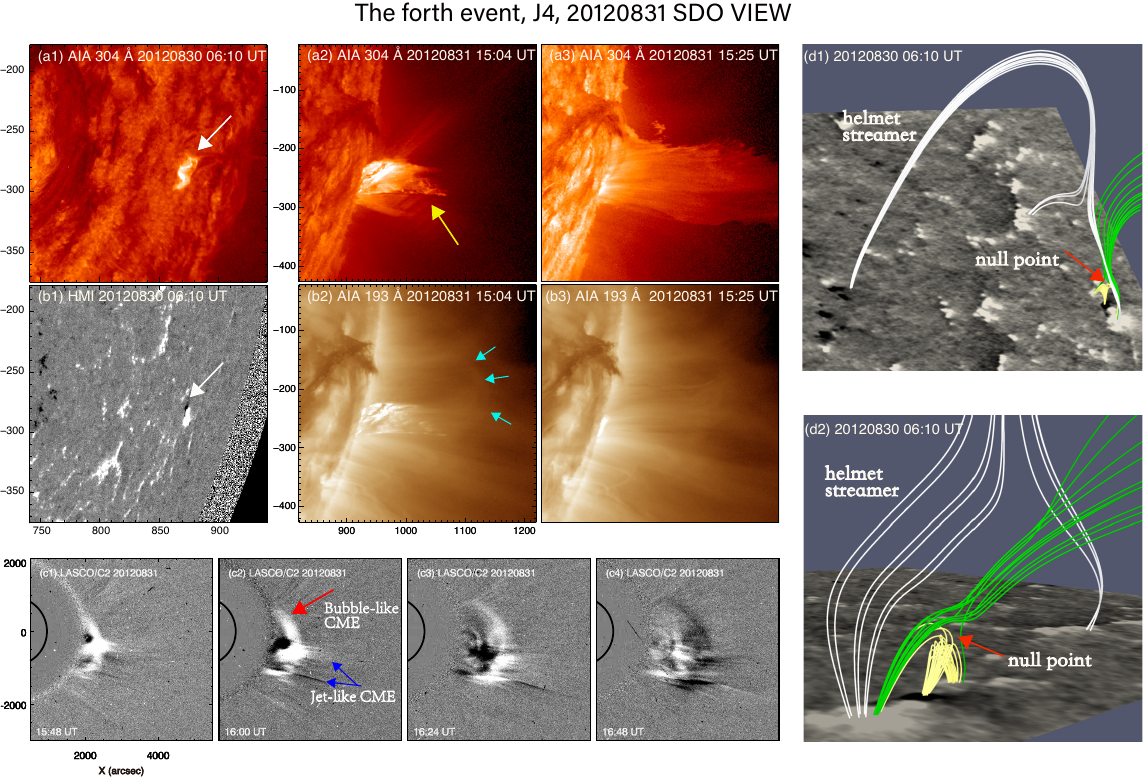}}
\caption{Photospheric magnetograms and EUV images from SDO, and the twin CME from LASCO, and potential field extrapolations for the source region of J4. (a1)$-$(b1) showing the source region one day before the jet eruption. The white arrows denote the eruptive region. The yellow arrow shows the jet and the cyan arrows point to the magnetic loop system around the jet eruption in (a2)$-$(b2). In (d1)$-$(d2), the extrapolated field lines overlying the HMI magnetograms, one day before the jet eruption. The green (yellow) lines represent the open (fan) field lines in panels (d1) and (d2). The red arrows point out the null point and the white lines represent the adjacent helmet streamer. An animation of the AIA 304 \AA sequence, panels (a2) - (a3) but with a larger field-of-view, is available from 14:44 UT to 15:25 UT. The duration of this animation is 9 s.}
\label{fig10}
\end{figure}

\subsection{The Kinematic Evolution of the J1$-$J4 Events and Associated Phenomena}
\par The kinematic evolution of the J1$-$J4, the associated Type \rmnum{3} radio bursts, and the kinematics of the two components of twin CMEs are shown in Figure 11. The time-distance diagrams of panels (a1)$-$(d1) are plotted along the main axes of J1-J4; for example, the time-distance diagram for J3 (c1) is plotted along the white arrow in Figure 12 panel (A). Based on the time-distance diagrams, we measured the speed of the rising filament and the ejected velocity of the jet. The relevant physical parameters are presented in the fifth and sixth columns of Table 1. In our 16 events, all jets are accompanied by Type \rmnum{3} radio bursts (refer to the fourth column of Table 1). The presence of Type \rmnum{3} radio bursts indicates the escape of near-relativistic electrons along open magnetic field lines into interplanetary space from the erupted source region. This implies that the twin CME and the narrow CME involve a process of magnetic reconnection between the source region and the open field magnetic field lines. We observed that the Type \rmnum{3} radio bursts associated with the eastern jets (e.g., J2) were weaker compared to those associated with the western jets (e.g., J3), despite the eastern jets (J2) being larger than the western ones (J3) (from the SDO viewpoint). This seems to be understandable based on the fact of Parker spirals. In addition, the directivity of Type \rmnum{3}s can be also influenced by the refraction effects of interplanetary density gradients and the scattering effects of random density fluctuations, as simulated by \citet{2007ApJ...671..894T}. In the case of our four twin CME events, the filament lifted earlier than the appearance of Type \rmnum{3} radio bursts. With the jets being ejected into the higher corona, two morphologies of CMEs (twin CMEs) in LASCO/C2 were captured. By combining observations from STEREO, the twin CMEs caused by a single jet were confirmed. Subsequently, the velocities of the twin CMEs were fitted, revealing that the speed of the jet-like component was higher than that of the bubble-like component.

\linespread{1.0}
\begin{table}[tb]\tiny \caption{Information about all Events}\label{t1}
\centering
\setlength{\tabcolsep}{0.3mm}{
\renewcommand\arraystretch{1}
\tabcolsep=0.01cm
\begin{tabular}{lccccccccccccc}
\hline
\hline
Twin CMEs  \\\hline
Jet & Date & Time &Type \rmnum{3} & V$_{J}$ & V$_{F}$  & T$_{JCME}$  &V$_{JCME}$   & T$_{BCME}$    & V$_{BCME}$ & filament &  Null point   \\
     &           &       & Time  &     &      &        & &    &  & length     & Height \\   
     &           & (UT)    & (UT) & $\speed$  &  $\speed $   &   (UT)  & $\speed$ & (UT) & $\speed$ &Mm       & Mm \\\hline     

J1  & 28 Aug 11    &04:15 &04:16  &  107. 35&35.84 & 05:00     &     637    & 04:48      & 355  &  50.93  &     40.3   \\
J2   & 20 Jan 12      & 21:57&21:52 &   340.18   & 28.73  & 22:36     &491    & 22:24       & 445 & 135.88 &  73.7           \\

J3   & 09 Aug 12      & 21:45&21:36  &   112.96   &27.91 &     21:45     & 439      &   21:45     & 388 &113.40 &   39.0/46.5        \\

J4   & 31 Agu 12      &15:00  &15:02&  120.90    &34.94& 15:36         & 588       & 15:36 &      243     &   54.23  &  61.9 \\
J5 \citet{2012ApJ...745..164S}   & 22 Jul 11      & 16:27  &16:23&   235   & 12.23& 16:45         &  309   & 16:55 &     $-$   &   52.29          &  21.0 \\
J6 \citet{2019SoPh..294...68S}   & 16 May 14      & 03:59 & 03:59 &   325    &11.76& 04:38         &    619 & 04:38 &      620   & 57.86       &  52.8   \\
J7 \citet{2019ApJ...881..132D}   & 23 Aug 15      & 04:20 &04:10 &   649   &3-8 & 04:36         &   895   &  05:00 &      340    &   56.50          & $-$  \\
\hline
Narrow CMEs   \\\hline
J8   & 1 Apr 11      &  03:54  &03:51&  363.93    &  unclear  &  04:17    &   541   & \dots &     \dots     &    21.03  & 3\\
J9   & 25 Aug 11      & 17:57 &17:57 &   110.21    & unclear &   19:01     &  220   & \dots &     \dots     &   18.06  &15.6 \\
J10   & 23 Dec 11      & 20:53  &20:52&   81.00    & unclear &    21:24    &   293    & \dots &     \dots     &    24.07     & 16.6\\
J11   & 11 May 12      & 13:55 &13:58 &   184.33    &  16.84 &  14:36     &    564  & \dots &     \dots      &     82.70 &44.6 \\
J12   & 11 Jun 12      &  16:00 &15:59&    83.25   &  9.156 & 16:36      &   357   & \dots &     \dots   &  29.39  & 22.4\\
J13   & 1 Aug 12      &  18:32 &18:28&    141.12   &  23.664 & 19:00   &  647   & \dots &     \dots &  54.23  & 38.2 \\
J14   & 14 Nov 12      & 22:39  &22:38&    132.41   & 8.677  &  23:12     &    556  & \dots &     \dots  &   50.72  &2.5 \\
J15   & 17 Nov 12      &  18:08 &18:08&   163.80    &   2.537 &  18:48    &  408   & \dots &     \dots  &  49.02  &5.1 \\
J16 \citet{2022ApJ...926L..39D} & 14 Jul 12      &  09:09 &09:08&   360-425    & 6.7 &  09:48     &    561   & \dots &     \dots   & 20.57          & 5.1\\

\hline
\end{tabular}
}
\\
\end{table}

\subsection{Narrow CMEs and Twin CMEs}
\par
To further investigate the factors contributing to the production of narrow CMEs or twin CMEs, various parameters were measured, including the length of the filament at the base of jets, the speed of jets, the twist number of jets, the null point of the fan-spine system, and the velocity of CMEs for all events, among other parameters. The detailed physical parameters can be found in Tables 1 and 2. Filament length was obtained using 3D reconstruction techniques \citep[e.g.,][]{2017ApJ...851..133Z,2019ApJ...883..104S,2021A&A...647A.112Z,2022MNRAS.516L..12T,2023MNRAS.520.3080T}. We utilized scc\_measure.pro \citep{2006A&A...449..791T} in the SSW package to manually obtain the 3D coordinates and image pixel positions of the filament features. The process involved the following steps: opening two 304 \AA~images from AIA and EUVI, selecting a feature in image A (AIA), identifying the corresponding filament structure in image B EUVI, and subsequently calculating and outputting the pixel positions of the structure in both images. The points used to project the filament structure traces are displayed in Figure 1 (a1)$-$(a4) and (b1)$-$(b4). It should be noted that most of the obtained filament lengths in Table 1 (J1$-$J16) were averaged between the measurements from AIA and STEREO views. For events where 3D reconstruction was not possible, filament lengths were derived from the AIA view.  By placing straight/curved lines along the main axis of jets in AIA 304 \AA~images and plotting time-distance diagrams, it is possible to estimate the speed of the jets. An example of such a curved line can be seen in Figure 12 (A) for the case of J3, while the corresponding time-distance display is shown in Figure 11 (c1).

\begin{figure} 
\centerline{\includegraphics[width=0.45\textwidth,clip=]{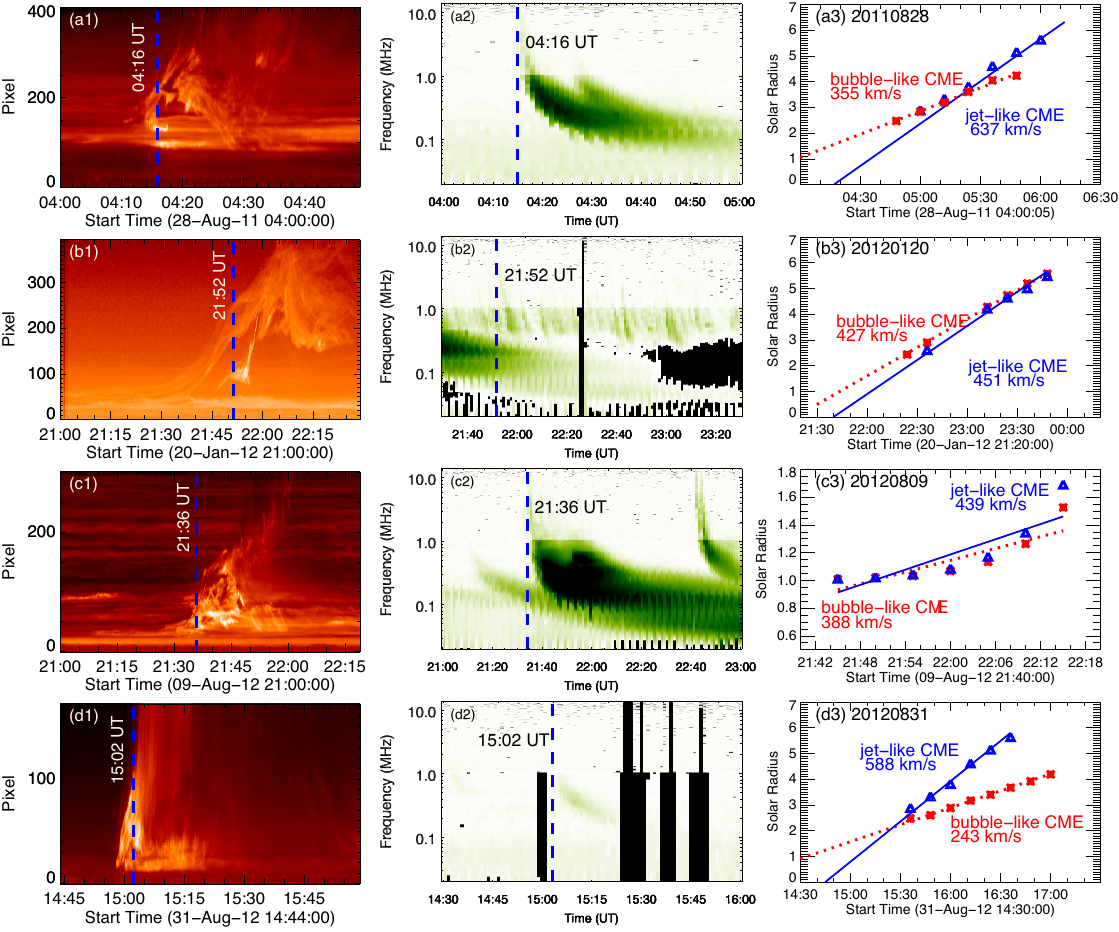}}
\caption{In (a1)$-$(d1), the time-distance intensity diagrams of AIA 304 \AA~ for J1-J4 are plotted along the main axis of each jet. The blue dotted lines correspond to the start time of Type \rmnum{3} radio bursts in (a2)$-$(d2). In (a3)$-$(d3), the measured data points (blue triangles and red asterisks) of fitting lines are all derived from tracing the front edge of the CMEs in LASCO/C2 (except (c3) from COR1).
}
\label{fig11}
\end{figure}

The twist contained under the fan dome is released onto the newly reconnected field lines, launching the rotational motion of an associated jet. The twist transferred to surrounding fields (open or closed fields) then helps drive the jet material outward away from the sun, forming the different morphology of CMEs. Following this scenario, we estimated the twist of the jet spire roughly by assuming that the main axis of the rotating jet was a cylinder and the helical structure was untwisted quickly following the method of \citet{2011ApJ...735L..43S}. A time-distance diagram is made for a slit cutting along the A-B direction (see example J3 in Figure 12 (c1)). The average rotational speed $\overline{v}_{r}$ of the untwisting jet (J3) is estimated to be \speed{91.92}, by tracing the bright stripes on the time-distance diagram and performing multiple linear fits (see Table 2 for the physical parameters of J3). The onset and cut-off times of the rotation of the bright stripes of the jet are shown in (c2), lasting for about 1560s. The average width (diameter) of the jet spire is 34.7 Mm as indicated by the green dashed line in (c2). The average angular velocity of the rotating plasma ($\omega=2\overline{v}_{r}/D$) is calculated to be 5.3$\times$10$^{-3}$ rad s$^{-1}$. The total amount of twist that the jet spire has released is roughly 1.3 turns or 2.6 $\pi$. The total number of turns of all 16 jets is calculated in a similar way as for J3. We plotted the time$-$varying intensity in a slit perpendicular to the jet body (such as Figure 12 (c1) and (c2)). The average rotational speed of the untwisted jet was estimated by tracing the bright stripes on the time-distance plot and performing multiple linear fits (such as white dashed lines in Figure 12 (c2) ). The average angular velocity of the rotating plasma ($\omega=2\overline{v}_{r}/D$) is calculated to obtain the number of twists released at the spire of this jet. The rotational speeds, the width, the angular velocities, 
the periods and the total numbers of turns of all jets are listed in Table 2, in which the jet rotations of J7, J10, J12, J14, and J16 were not measurable.

\linespread{1.0}
\begin{table}[tb]\tiny \caption{Information about all Events} \label{t1}
\centering
\setlength{\tabcolsep}{0.3mm}{
\renewcommand\arraystretch{1}
\tabcolsep=0.01cm
\begin{tabular}{lcccccccccc}
\hline

Twin CMEs  \\\hline
Jet & Date & Time &Dur & $\overline{V}$$_{r}$ & Diameter  & $\omega$  &Period   &Twist    & Twist    \\
     &         &   (UT)    &  (s)   & $\speed$  &( Mm )   & (10$^{-3}$rad s$^{-1}$)&(s) &(turn)  &($\pi$) \\   
\hline
J1  & 28 Aug 11    &04:15 & 1800 &78.1&42.0&3.7&1690& 1.1&2.2\\
J2   & 20 Jan 12      & 21:57&1980& 118.6& 43.2&5.5&1143&1.7&3.4         \\
J3   & 09 Aug 12      & 21:45& 1560 &91.9 & 34.7& 5.3&1187&1.3&2.6\\
J4   & 31 Agu 12      &15:00  &1380 &127.7&63.5&4.0&1562&0.9&1.8   \\
J5 \citet{2012ApJ...745..164S}   & 22 Jul 11      & 16:27  &540& 64.8&20.9&6.2&1014&0.5& 1.0  \\
J6 \citet{2019SoPh..294...68S}   & 16 May 14    &03:59 & 1560 &111.2&39.2&5.7&1107&1.4&2.8\\
J7 \citet{2019ApJ...881..132D}   & 23 Aug 15      & 04:20 &\dots &    \\
\hline
Narrow CMEs   \\\hline
J8   & 1 Apr 11      &  03:54  &960&126.3&18.7&13.5&464&2.0& 4.0 \\
J9   & 25 Aug 11      & 17:57 & 1620&41.8&17.9&4.7&1343 &1.2&2.4      \\
J10   & 23 Dec 11      & 20:53  &\dots& \\
J11   & 11 May 12      & 13:55 &300&105.4&20.7&10.2&618& 0.5&1.0  \\
J12   & 11 Jun 12      &  16:00 &\dots&  \\
J13   & 1 Aug 12      &  18:32 &1140&116.2&56.3&4.1&1522&0.7&1.4   \\
J14   & 14 Nov 12      & 22:39  &\dots&    \\
J15   & 17 Nov 12      &  18:08 &960&100.5&22.0&9.1&687&1.4&2.8 \\
J16 \citet{2022ApJ...926L..39D} & 14 Jul 12      &  09:09 &\dots&  \\

\hline
\end{tabular}
}
\\
\footnotesize{\textbf{Note.} $\overline{v}_{r}$, Diameter, $\omega$, period indicate the average rotational speed of jets, the width of the jet spire, the angular speed and period of the jet, respectively.}\\
\end{table}

\begin{figure}
\centerline{\includegraphics[width=0.45\textwidth,clip=]{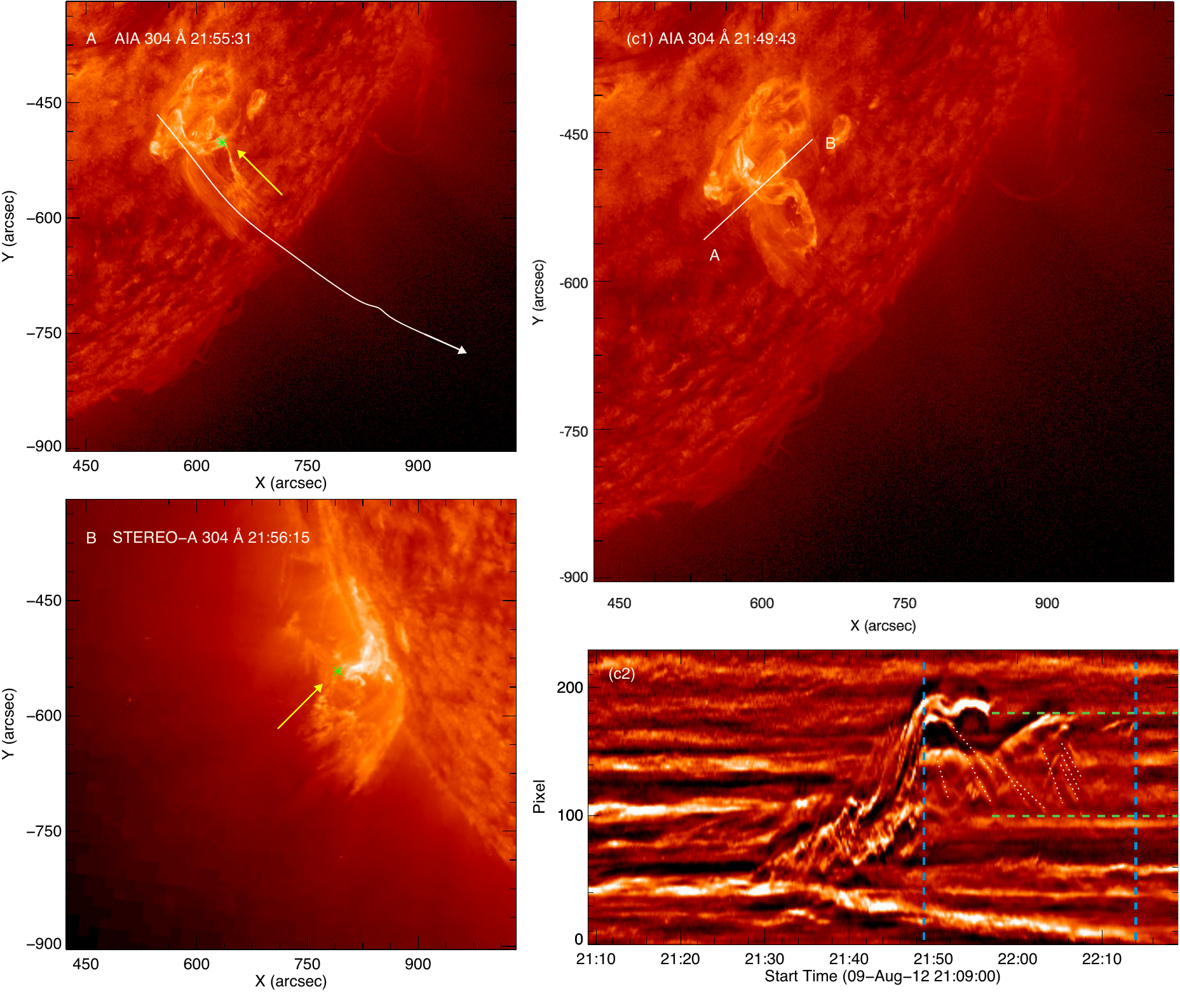}}
\caption{The example is from 2012 August 9 (J3) to show how we can estimate the reconnection height and the twist of the jet spire. In panels A and B, the green asterisks pointed by the yellow arrows represent the null point, obtained by three-dimensional reconstruction. The white arrow marks the slit position of the time-distance plot shown in Figure 11 (c1). (c2) the time-distance intensity plot along the white line A$-$B in (c1). The blue dotted lines serve as the beginning and end times of the rotating jet. The width of the rotating jet is described by two green dotted lines.}
\label{fig12}
\end{figure}

Furthermore, taking advantage of the PFSS magnetic extrapolation technology and local provided by the SSW, we also estimated the null point height of the fan-spine topology of all 16 jets. First, 300 magnetic field lines were randomly selected in the eruptive source region, and the approximate positions of the fan-spine structure were identified. Subsequently, the focus was on the height of the inner spine of the fan-spine configuration, and through multiple attempts, the height just before the disappearance of the inner spine was selected as the location of the magnetic null point. Figure 13 (a1)$-$(b1) show the three-dimensional (3D) coronal magnetic field structures of the J3 and J16 eruptions beforehand, respectively, obtained by extrapolation from PFSS. The relevant extrapolated magnetic lines are superimposed on the AIA 304 \AA~ images, as shown in Figure 13 (a1)$-$(b1). It is clear that the structures are the fan-spine configuration, which consists of the dome fan (represented by yellow lines) and the field lines close to the outer spine (represented by green lines). The white lines correspond to the nearby closed field lines (helmet streamer). By extrapolating the magnetic field, we obtained the height of the null point of J3 (J16) at approximately 39 (5.1) Mm. Although the null point heights of most events (J1$-$J16) can be obtained through magnetic field extrapolation techniques, not all jets can be extrapolated due to limitations in the available magnetic field data. In such cases, the height of the null point was approximated using scc\_measure.pro based on 3D reconstruction. Our criterion for determining the reconnection height was the presence of material propagation along open field lines, while both ends of the filament's foot points remained rooted in the solar surface (see the green asterisks in the example of J3 in Figure 12 (A)$-$(B)). Using this 3D reconstruction, the initial null height of J3 was determined to be 46.5 Mm. Although there is a margin of error in obtaining the null point height between 3D reconstruction and the PFSS method, the error remains within an acceptable range. 

\begin{figure}
\centerline{\includegraphics[width=0.45\textwidth,clip=]{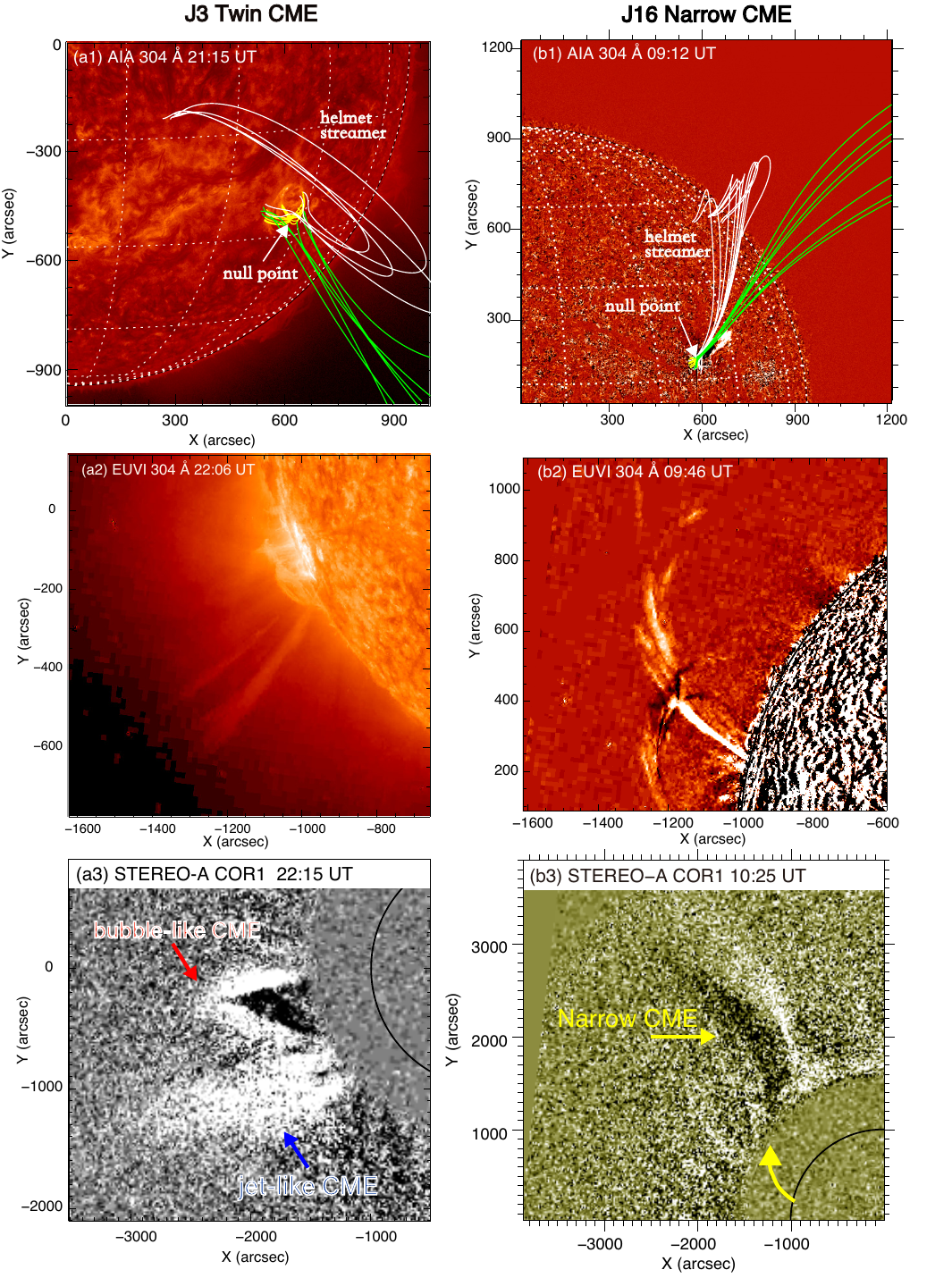}}
\caption{Comparison between jet-related twin CME and jet-related narrow CME in an initial analogous magnetic topology. (a1)$-$(a3) display the twin CME from the event J3 on 2012 August 9, while (b1)$-$(b3) depict the narrow CME on 2012 July 14 \citep{2022ApJ...926L..39D}. Panels (b1)–(b2) are the running-difference images. From the SDO/AIA perspective, (a1)$-$(b1) illustrates these jets occurring under the fan-spine topology, both are located near a helmet streamer system. The corresponding STEREO/EUVI viewpoint of the same jet's morphology is displayed in (a2)-(b2). In (a1)$-$(b1), the yellow (white and green) lines represent the closed fan dome and the field lines around the outer spine. (a3) and (b3) show the CME morphology in different events as observed by STEREO/COR1. The yellow arrow marks the ejected direction of the jet in panel (b3).}
\label{fig13}
\end{figure}

The scatter plot analyses of various physical parameters provide insights into the factors influencing the production of twin CMEs versus narrow CMEs, considering their similar topological structures characterized by magnetic reconnections between filaments and surrounding magnetic fields. The results are shown in Figure 14. Some statistical works from observations showed a longer filament tends to reside higher in the corona \citep[e.g.,][]{2018ApJ...857L..14X}. Therefore, the dome of a fan-spine structure should have a larger size and a higher null-point height to hold or confine a longer filament prior to its eruption. Based on this assumption, we conducted a fitting analysis to establish the linear relationship between the length of the filament and the null point height of the fan-spine system. Notably, it is intriguing that in 83\% of the observed twin CMEs, the null point height exceeds 40 Mm, as depicted by the green line in Figure 14 (a). Furthermore, in 100\% of twin CME events, the length of the filament associated with the eruptions from the source region slightly exceeds 50 Mm, as indicated by the pink vertical line in Figure 14(a). The correlation between the length of the filament and the null point height is approximately 70\% in our study. 

\begin{figure}
\centerline{\includegraphics[width=0.45\textwidth,clip=]{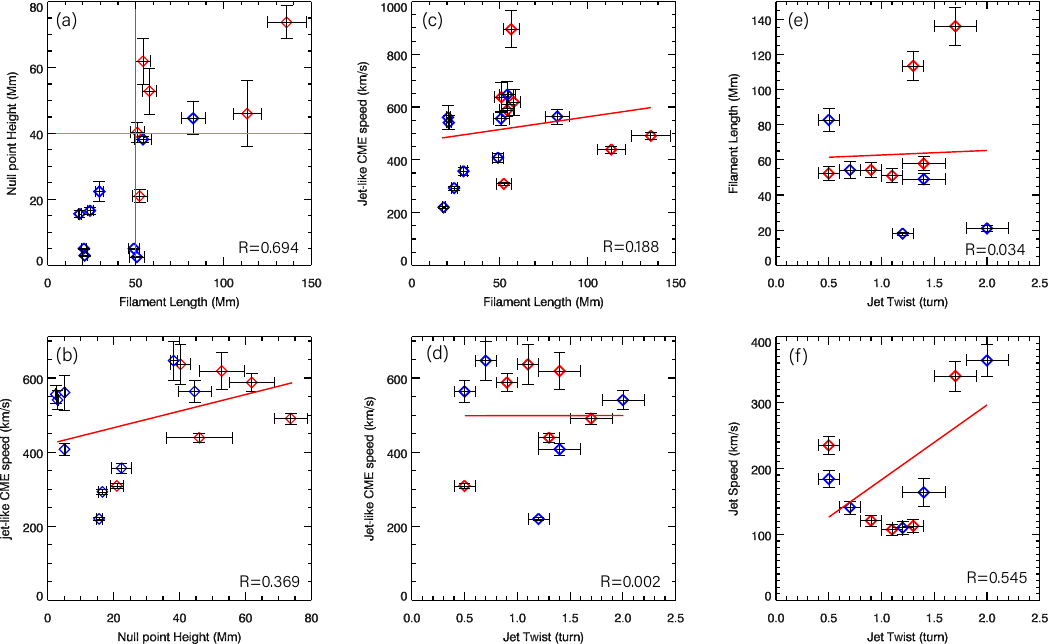}}
\caption{Parameter comparison between narrow CMEs and twin CMEs. The blue crosses represent the jet-related narrow CMEs J8$-$J16, while the red crosses express the jet-related twin CMEs J1$-$J7. Panel (a): the pink line represents the filament length of 50 Mm, while the green indicates the null point height of 40 Mm. The error of these values is obtained by calculating the difference between the maximum and minimum values obtained from multiple measurements of the average value.}
\label{fig14}
\end{figure}

In terms of the correlation strength from high to low, the following relationships are observed: (1) The twist of the jets shows a moderate correlation (approximately 54\%) with the velocity of the jets themselves. (2) The null point height demonstrates a relatively weaker correlation (approximately 37\%) with the velocity of jet-like CMEs. It is important to note that the term ''jet-like CME speed'' refers specifically to the velocity measured both from narrow jet CMEs and the jet-like component of twin CMEs. According to the breakout eruption model \citep{1999ApJ...510..485A, 2017Natur.544..452W}, the null point does not remain a null once the filament-supporting flux begins to rise and the fan dome expands. The magnetic pressure distorts the null into a current sheet, which enables reconnection to occur in several locations (and plasmoids to form). After the flare reconnection forms the flux rope, the upper magnetic reconnection site migrates \citep[e.g.,][]{2017Natur.544..452W,2018ApJ...852...98W}. The point is that the initial null height is not necessarily the height at which the filament-carrying flux reconnects with external flux, so it’s no surprise that the height correlation is poor. (3) The length of the filament exhibits a weaker correlation (approximately 19\%) with the velocity of jet-like CMEs. On the other hand, the following relationships show no apparent correlation: (4) The jet twist and the length of the filament have a negligible correlation (approximately 3\%). (5) The jet twist and the speed of jet-like CMEs have a minimal correlation (approximately 0.2\%).

\section{Summary and Discussion} \label{sec:summ}

Combined with the SDO/AIA and STEREO/EUVI two perspectives, we performed a statistical study of jet-related twin CMEs and jet-related narrow CMEs from Oct 2010 to Dec 2012. Our study revealed that all CMEs in our sample are accompanied by filament-driven jets and Type \rmnum{3} radio bursts during their initial formation and involve magnetic reconnection between twists contained in the fan dome and surrounding magnetic fields, and most events occur within a fan-spine system. We explored the origins of the two components of twin CMEs. Furthermore, we examined the factors influencing the occurrence of narrow CMEs and twin CMEs, encompassing the measurement of various parameters and conducting a comprehensive correlation analysis. Our study suggests that part of the twists contained under the 3D fan-spine magnetic field configuration transfers onto open fields and then propels the jet to propagate into interplanetary space, manifesting as a jet-like CME. When a significant portion of the twist is injected into the neighboring closed fields (helmet streamer), the closed fields expand outward and manifest as a bubble-like CME in coronagraph images. Moreover, our findings suggest that the length of filaments and the initial null point height may serve as potential physical factors in determining the morphology of jet-related CMEs in the high corona. Our observational results reveal the following: 1) 83\% of jet-related twin CME occurrences take place when the length of the filament at the base of the jet is longer than 50 Mm and the initial null height is higher than 40 Mm; conversely, 89\% of the jet-narrow CME events occur with a filament smaller than 50 Mm and the null point height below 40 Mm.

\par
This article analyzes 16 jet events characterized by jet-CMEs accompanied by Type \rmnum{3} solar radio bursts. Among these events, 7 are associated with twin CMEs (jet-like and bubble-like components), while 9 are only associated with narrow CMEs. Moreover, all of the twists contained in the fan dome changed to the surrounding fields, suggesting the occurrence of the magnetic reconnection process. The velocities of the jets are in the range of \speed{81$-$649}, which is within the range of \speed{10$-$1000} obtained for 100 X-ray jets in the previous work of \cite{2000ApJ...542.1100S} and similar to the velocities of \speed{87$-$532} obtained by \cite{2016A&A...589A..79M} for 20 EUV jets. The velocity of the filament rise is \speed{3$-$35}. The velocities are \speed{293$-$895} for the narrow CMEs (and/or jet-like component), and \speed{243$-$723} for bubble-like CMEs. The filament lengths are 20$-$135 Mm and the null point height ranges from 2.5$-$73.7 Mm. The duration time of the jet is 540$-$1980 s, the angular velocity is 3.7$-$13.5$\times$10$^{-3}$ rad/s, the released twist of the jet spire is about 0.5$-$2 turns, 1$-$4 $\pi$, similar with previous works \citep[e.g.,][]{2011ApJ...735L..43S,2018ApJ...852...10L,2019ApJ...871..229C,2019FrASS...6...44L,2021ApJ...911...33C}.

\par
There is evidence that both narrow and bubble types of CME may be driven by the jets. The twist at the base of the jet can be transferred to the ambient open fields, propagating into interplanetary space and forming the narrow CME. On the other hand, the twist can also be transferred to the surrounding closed loops, enabling them to expand outward and form the bubble-like CME. \citep[e.g.,][]{2016ApJ...822L..23P,2021ApJ...911...33C}. Regarding whether a bubble-like or narrow CME is produced under the fan-spine topology, \cite{2021ApJ...907...41K} contended that it is related to the ratio of the magnetic free energy in the filament channel to the free energy of the field inside and outside the overlying dome. What physical factors determine the morphology of jet-related CMEs in the high corona? In the current work, we have conducted further research on the magnetic field topology during the occurrence of jets that cause twin CMEs. We have also compared the similarities and differences in some physical parameters between narrow CMEs and twin CMEs caused by jets. 
\par
It is worth noting that Figure 13 displays the similar magnetic field configurations of two jets occurring within the three-dimensional fan-spine framework, which reside on the flank of the closed fields. This configuration is consistent with the simulations conducted by \cite{2021ApJ...909...54W}. In their simulation, initially, the external reconnection (or breakout reconnection) would distort the null point into a current sheet (breakout current sheet). Reconnection at this current sheet then weakens the overlying fields above the filament, enabling the filament to rise \citep[e.g.,][]{2017Natur.544..452W,2018ApJ...852...98W}. An eruptive change in evolution occurs as the filament reaches the current sheet, after which an untwisting jet is launched. When the jet source is adjacent to a helmet streamer, the enhanced breakout reconnection above the filament channel also causes the twisted jet magnetic field to enter the surrounding fields, partially along the closed flux beneath the helmet streamer and blowing off the top of the streamer, resulting in a bubble-like CME. However, what sets J3 and J16 apart is that jet-related CMEs exhibit different morphologies in the high solar corona. One is a narrow-band CME (Figure 13 (b3)), while the other is a twin CME (Figure 13 (a3)) composed of both a jet-like component and a bubble-like structure. Based on the statistical results, the length of filaments and the initial null point height may be the possible physical factors for determining the morphology of jet-related CMEs in the high corona. 

\begin{figure}
\centerline{\includegraphics[width=0.45\textwidth,clip=]{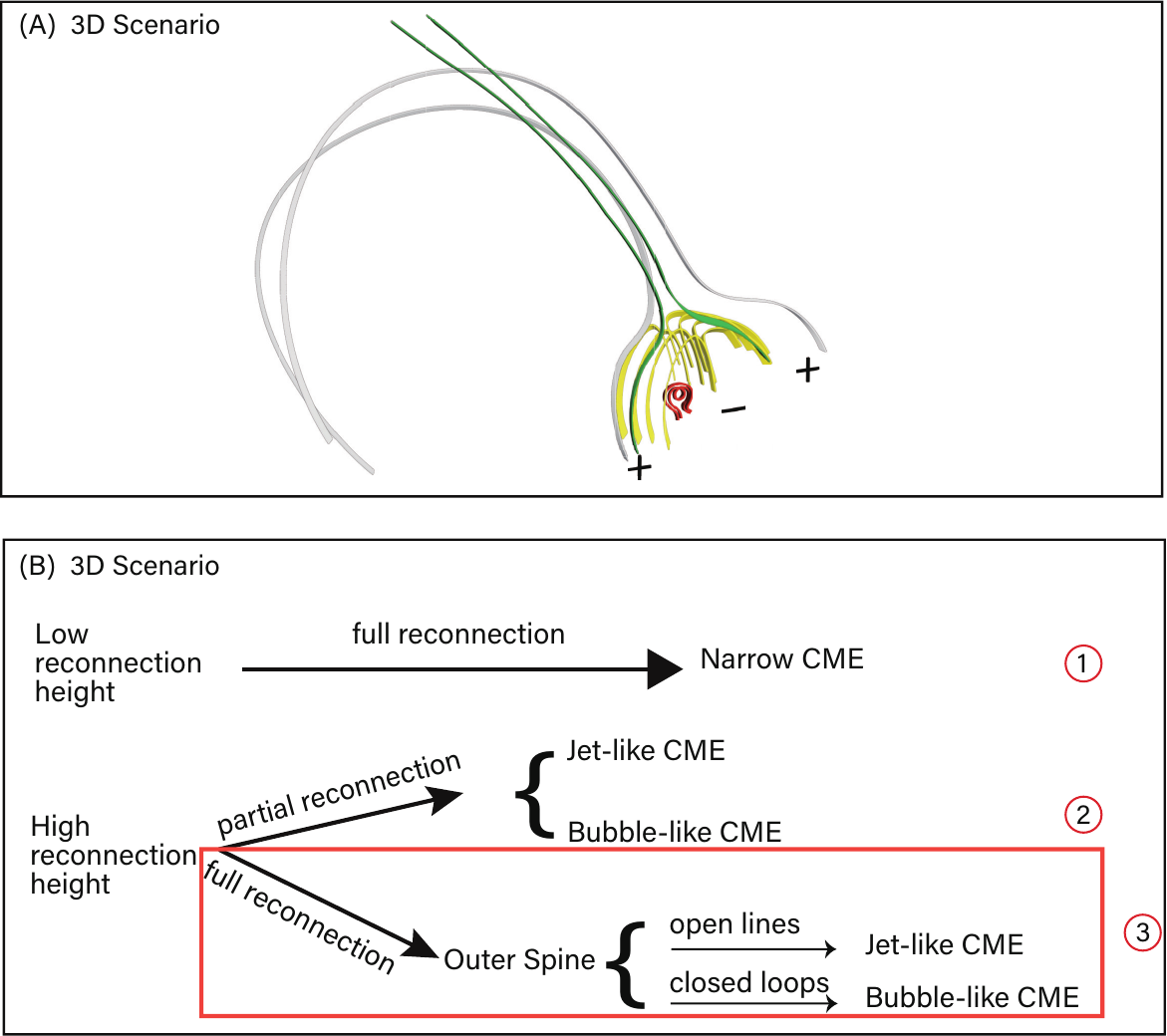}}
\caption{Schematic indicates the magnetic field topology of the source region for the twin CMEs in our study. In (A), the open fields are depicted by the green lines and closed fields (helmet streamer) adjacent to the fan dome (yellow surface) are represented by the white lines. (B) illustrates the several possible scenarios for the partial eruption and full eruption of the closed magnetic flux under the fan dome at different heights of the magnetic null point. }
\label{fig15}
\end{figure}

This twin CME phenomenon was first reported by \cite{2012ApJ...745..164S} who suggested that the jet-like component is generated by the external reconnection between the field overlying the rising filament and the surrounding open field, and the bubble-like CME is caused by the filament eruption. In the process of external reconnection, they considered that the flux of the filament was not completely destroyed. \citet{2019ApJ...881..132D} further fit the velocity of the twin CME by extrapolating the height of the twin CME back to the solar surface and revealed that the initiation of the jet-like CME coincides approximately with the onset time of the jet, while the initiation of the bubble-like CME aligns approximately with the time of the filament eruption. In their work, a jet-like CME was first observed by LASCO, despite the filament erupting first. They explain that a jet-like CME may have an acceleration process. This acceleration of the jet may be due to the slingshot effect of the bent open fields induced by the interaction. The acceleration of the jet may also be attributed to the untwisting of the jet, which signifies the release of non-potential magnetic energy stored in the pre-eruption filament. In contrast, \cite{2020ApJ...893..115C} suggests that the twin CME caused by the jet is due to the bifurcation of the jet spire. In our study, we suggest that our results complement the explanation of \cite{2021ApJ...907...41K}. In their work, they contended that a critical quantity for determining the type of eruption (jet-like component or bubble-like component) is the amount of free magnetic energy stored in the sheared filament channel. In this current work, the length of the filaments may serve as an indicator of the magnetic free energy associated with the filament channel to some extent. In the 3D fan-spine scenario, the smaller the filament, the lower the null-point height, implying that the twisted magnetic flux contained within the closed magnetic flux system could more easily be fully transferred to the open field through effective magnetic reconnection, resulting in a narrow CME, as illustrated by case 1 in Figure 15 (B). On the other hand, if the filament is longer and the reconnection height is higher, only a partial transfer of the closed magnetic flux to the open field occurs, forming a jet-like component of a twin CME. A partial eruption of the closed magnetic flux carrying the filament directly leads to the formation of a bubble-like CME in the twin CME scenario. This scenario is consistent with the twin CME formation mechanism proposed by \cite{2012ApJ...745..164S}, as illustrated by case 2 in Figure 15 (B). However, in this paper, we discussed a third scenario where the fan-spine structure is located near the helmet streamer, which is consistent with the simulations conducted by \citet{2021ApJ...909...54W}, as shown in Figure 15 (A) and the red box in (B). 
\par
In our study, the field carrying the filament (as depicted by the red lines in Figure 15 (A)) reconnects with and releases twists onto the open fields, producing a jet. The jet propagates into interplanetary space exhibiting a jet-like CME. In addition, a significant portion of the twist is injected into the closed fields beneath the adjacent helmet streamer, leading to the outward expansion of the closed fields and the formation of a bubble-like CME. The height of the null point also governs the available time for the build-up of the filament (or flux rope), once the flare reconnection commences and the filament (or flux rope) ascends towards the breakout current sheet. This affects whether a partial or complete of the twisted flux can be transferred onto external open field lines, and hence whether a jet or a jet-like CME is expelled. On the other hand, assuming that the length of the filament is proportional to the sheared filament channel (which may not necessarily be accurate if there is no filament mass accumulation within the channel), more filament length adds more free energy. Therefore, the combination of null point height and the length of the filament could reinforce each other. In other words, it appears that in cases where the initial null point is higher and the length of the filament under the fan-spine system is longer, there may be a stronger driving force for the expansion of the surrounding closed fields. This may suggest that driving a bubble-like component of a twin CME requires a greater amount of magnetic energy stored under the fan dome. Of course, more observations and simulation work are needed to confirm the formation mechanism of twin CMEs.

The authors are grateful for the anonymous referee’s valuable comments and suggestions. The authors are grateful for the data provided by the SDO, SOHO/LASCO, and STEREO science teams and helpful discussions with Dr. Hechao Chen from Yunnan University. This work is supported by the Beijing Natural Science Foundation (1244053) and National Postdoctoral Programs (GZC20230097, 2023M740112). Y.D.S. was supported by the Natural Science Foundation of China (12173083, 11922307), the Yunnan Science Foundation for Distinguished Young Scholars (202101AV070004), and the National Key R\&D Program of China (2019YFA0405000).


\begin{thebibliography}{}
\expandafter\ifx\csname natexlab\endcsname\relax\def\natexlab#1{#1}\fi
\providecommand{\url}[1]{\href{#1}{#1}}
\providecommand{\dodoi}[1]{doi:~\href{http://doi.org/#1}{\nolinkurl{#1}}}
\providecommand{\doeprint}[1]{\href{http://ascl.net/#1}{\nolinkurl{http://ascl.net/#1}}}
\providecommand{\doarXiv}[1]{\href{https://arxiv.org/abs/#1}{\nolinkurl{https://arxiv.org/abs/#1}}}

\bibitem[{{Adams} {et~al.}(2014){Adams}, {Sterling}, {Moore}, \&
  {Gary}}]{2014ApJ...783...11A}
{Adams}, M., {Sterling}, A.~C., {Moore}, R.~L., \& {Gary}, G.~A. 2014, \apj,
  783, 11, \dodoi{10.1088/0004-637X/783/1/11}

\bibitem[{{Antiochos} {et~al.}(1999){Antiochos}, {DeVore}, \&
  {Klimchuk}}]{1999ApJ...510..485A}
{Antiochos}, S.~K., {DeVore}, C.~R., \& {Klimchuk}, J.~A. 1999, \apj, 510, 485,
  \dodoi{10.1086/306563}

\bibitem[{{Bougeret} {et~al.}(1995){Bougeret}, {Kaiser}, {Kellogg}, {Manning},
  {Goetz}, {Monson}, {Monge}, {Friel}, {Meetre}, {Perche}, {Sitruk}, \&
  {Hoang}}]{1995SSRv...71..231B}
{Bougeret}, J.~L., {Kaiser}, M.~L., {Kellogg}, P.~J., {et~al.} 1995, \ssr, 71,
  231, \dodoi{10.1007/BF00751331}

\bibitem[{{Brueckner} {et~al.}(1995){Brueckner}, {Howard}, {Koomen},
  {Korendyke}, {Michels}, {Moses}, {Socker}, {Dere}, {Lamy}, {Llebaria},
  {Bout}, {Schwenn}, {Simnett}, {Bedford}, \& {Eyles}}]{1995SoPh..162..357B}
{Brueckner}, G.~E., {Howard}, R.~A., {Koomen}, M.~J., {et~al.} 1995, \solphys,
  162, 357, \dodoi{10.1007/BF00733434}

\bibitem[{{Carley} {et~al.}(2012){Carley}, {McAteer}, \&
  {Gallagher}}]{2012ApJ...752...36C}
{Carley}, E.~P., {McAteer}, R.~T.~J., \& {Gallagher}, P.~T. 2012, \apj, 752,
  36, \dodoi{10.1088/0004-637X/752/1/36}

\bibitem[{{Chen} {et~al.}(2013){Chen}, {Bastian}, {White}, {Gary}, {Perley},
  {Rupen}, \& {Carlson}}]{2013ApJ...763L..21C}
{Chen}, B., {Bastian}, T.~S., {White}, S.~M., {et~al.} 2013, \apjl, 763, L21,
  \dodoi{10.1088/2041-8205/763/1/L21}

\bibitem[{{Chen} {et~al.}(2018){Chen}, {Duan}, {Yang}, {Yang}, \&
  {Dai}}]{2018ApJ...869...78C}
{Chen}, H., {Duan}, Y., {Yang}, J., {Yang}, B., \& {Dai}, J. 2018, \apj, 869,
  78, \dodoi{10.3847/1538-4357/aaead1}

\bibitem[{{Chen} {et~al.}(2020){Chen}, {Hong}, {Yang}, {Xu}, \&
  {Yang}}]{2020ApJ...902....8C}
{Chen}, H., {Hong}, J., {Yang}, B., {Xu}, Z., \& {Yang}, J. 2020, \apj, 902, 8,
  \dodoi{10.3847/1538-4357/abb1c1}

\bibitem[{{Chen} {et~al.}(2021){Chen}, {Yang}, {Hong}, {Li}, \&
  {Duan}}]{2021ApJ...911...33C}
{Chen}, H., {Yang}, J., {Hong}, J., {Li}, H., \& {Duan}, Y. 2021, \apj, 911,
  33, \dodoi{10.3847/1538-4357/abe6a8}

\bibitem[{{Chen} {et~al.}(2019){Chen}, {Zheng}, {Li}, {Ma}, {Bi}, \&
  {Yang}}]{2019ApJ...871..229C}
{Chen}, H., {Zheng}, R., {Li}, L., {et~al.} 2019, \apj, 871, 229,
  \dodoi{10.3847/1538-4357/aafa83}

\bibitem[{{Chen} {et~al.}(2015){Chen}, {Su}, {Yin}, {Priya}, {Zhang}, {Liu},
  {Xu}, \& {Yu}}]{2015ApJ...815...71C}
{Chen}, J., {Su}, J., {Yin}, Z., {et~al.} 2015, \apj, 815, 71,
  \dodoi{10.1088/0004-637X/815/1/71}

\bibitem[{{Chen}(2011)}]{2011LRSP....8....1C}
{Chen}, P.~F. 2011, Living Reviews in Solar Physics, 8, 1,
  \dodoi{10.12942/lrsp-2011-1}

\bibitem[{{Chrysaphi} {et~al.}(2020){Chrysaphi}, {Reid}, \&
  {Kontar}}]{2020ApJ...893..115C}
{Chrysaphi}, N., {Reid}, H. A.~S., \& {Kontar}, E.~P. 2020, \apj, 893, 115,
  \dodoi{10.3847/1538-4357/ab80c1}

\bibitem[{{Duan} {et~al.}(2019){Duan}, {Shen}, {Chen}, \&
  {Liang}}]{2019ApJ...881..132D}
{Duan}, Y., {Shen}, Y., {Chen}, H., \& {Liang}, H. 2019, \apj, 881, 132,
  \dodoi{10.3847/1538-4357/ab32e9}

\bibitem[{{Duan} {et~al.}(2022){Duan}, {Shen}, {Zhou}, {Tang}, {Zhou}, \&
  {Tan}}]{2022ApJ...926L..39D}
{Duan}, Y., {Shen}, Y., {Zhou}, X., {et~al.} 2022, \apjl, 926, L39,
  \dodoi{10.3847/2041-8213/ac4df2}

\bibitem[{{Duan} {et~al.}(2024){Duan}, {Tian}, {Chen}, {Shen}, {Sun}, {Hou}, \&
  {Li}}]{2024ApJ...962L..38D}
{Duan}, Y., {Tian}, H., {Chen}, H., {et~al.} 2024, \apjl, 962, L38,
  \dodoi{10.3847/2041-8213/ad24f3}

\bibitem[{{Freeland} \& {Handy}(1998)}]{1998SoPh..182..497F}
{Freeland}, S.~L., \& {Handy}, B.~N. 1998, \solphys, 182, 497,
  \dodoi{10.1023/A:1005038224881}

\bibitem[{{Gopalswamy} {et~al.}(2014){Gopalswamy}, {Akiyama}, {Yashiro}, {Xie},
  {M{\"a}kel{\"a}}, \& {Michalek}}]{2014GeoRL..41.2673G}
{Gopalswamy}, N., {Akiyama}, S., {Yashiro}, S., {et~al.} 2014, \grl, 41, 2673,
  \dodoi{10.1002/2014GL059858}

\bibitem[{{Hong} {et~al.}(2017){Hong}, {Jiang}, {Yang}, {Li}, \&
  {Xu}}]{2017ApJ...835...35H}
{Hong}, J., {Jiang}, Y., {Yang}, J., {Li}, H., \& {Xu}, Z. 2017, \apj, 835, 35,
  \dodoi{10.3847/1538-4357/835/1/35}

\bibitem[{{Hong} {et~al.}(2011){Hong}, {Jiang}, {Zheng}, {Yang}, {Bi}, \&
  {Yang}}]{2011ApJ...738L..20H}
{Hong}, J., {Jiang}, Y., {Zheng}, R., {et~al.} 2011, \apjl, 738, L20,
  \dodoi{10.1088/2041-8205/738/2/L20}

\bibitem[{{Hou} {et~al.}(2019){Hou}, {Li}, {Yang}, \&
  {Zhang}}]{2019ApJ...871....4H}
{Hou}, Y., {Li}, T., {Yang}, S., \& {Zhang}, J. 2019, \apj, 871, 4,
  \dodoi{10.3847/1538-4357/aaf4f4}

\bibitem[{{Hou} {et~al.}(2023){Hou}, {Tian}, {Su}, {Madjarska}, {Chen},
  {Zheng}, {Bai}, \& {Deng}}]{2023ApJ...953..171H}
{Hou}, Z., {Tian}, H., {Su}, W., {et~al.} 2023, \apj, 953, 171,
  \dodoi{10.3847/1538-4357/ace31b}

\bibitem[{{Huang} {et~al.}(2024){Huang}, {Tan}, {Zhang}, {Zhu}, {Yang}, \&
  {Deng}}]{2024ApJ...965..137H}
{Huang}, J., {Tan}, B., {Zhang}, Y., {et~al.} 2024, \apj, 965, 137,
  \dodoi{10.3847/1538-4357/ad3353}

\bibitem[{{Huang} {et~al.}(2018){Huang}, {Xia}, {Nelson}, {Liu}, {Wiegelmann},
  {Tian}, {Klimchuk}, {Chen}, \& {Li}}]{2018ApJ...854...80H}
{Huang}, Z., {Xia}, L., {Nelson}, C.~J., {et~al.} 2018, \apj, 854, 80,
  \dodoi{10.3847/1538-4357/aaa9ba}

\bibitem[{{Illing} \& {Hundhausen}(1985)}]{1985JGR....90..275I}
{Illing}, R.~M.~E., \& {Hundhausen}, A.~J. 1985, \jgr, 90, 275,
  \dodoi{10.1029/JA090iA01p00275}

\bibitem[{{Innes} {et~al.}(2011){Innes}, {Cameron}, \&
  {Solanki}}]{2011A&A...531L..13I}
{Innes}, D.~E., {Cameron}, R.~H., \& {Solanki}, S.~K. 2011, \aap, 531, L13,
  \dodoi{10.1051/0004-6361/201117255}

\bibitem[{{Kaiser} {et~al.}(2008){Kaiser}, {Kucera}, {Davila}, {St. Cyr},
  {Guhathakurta}, \& {Christian}}]{2008SSRv..136....5K}
{Kaiser}, M.~L., {Kucera}, T.~A., {Davila}, J.~M., {et~al.} 2008, \ssr, 136, 5,
  \dodoi{10.1007/s11214-007-9277-0}

\bibitem[{{Karpen} {et~al.}(2017){Karpen}, {DeVore}, {Antiochos}, \&
  {Pariat}}]{2017ApJ...834...62K}
{Karpen}, J.~T., {DeVore}, C.~R., {Antiochos}, S.~K., \& {Pariat}, E. 2017,
  \apj, 834, 62, \dodoi{10.3847/1538-4357/834/1/62}

\bibitem[{{Klassen} {et~al.}(2011){Klassen}, {G{\'o}mez-Herrero}, \&
  {Heber}}]{2011SoPh..273..413K}
{Klassen}, A., {G{\'o}mez-Herrero}, R., \& {Heber}, B. 2011, \solphys, 273,
  413, \dodoi{10.1007/s11207-011-9735-4}

\bibitem[{{Kumar} {et~al.}(2018){Kumar}, {Karpen}, {Antiochos}, {Wyper},
  {DeVore}, \& {DeForest}}]{2018ApJ...854..155K}
{Kumar}, P., {Karpen}, J.~T., {Antiochos}, S.~K., {et~al.} 2018, \apj, 854,
  155, \dodoi{10.3847/1538-4357/aaab4f}

\bibitem[{{Kumar} {et~al.}(2019){Kumar}, {Karpen}, {Antiochos}, {Wyper},
  {DeVore}, \& {DeForest}}]{2019ApJ...873...93K}
{Kumar}, P., {Karpen}, J.~T., {Antiochos}, S.~K., {et~al.} 2019, \apj, 873, 93, \dodoi{10.3847/1538-4357/ab04af}

\bibitem[{{Kumar} {et~al.}(2021){Kumar}, {Karpen}, {Antiochos}, {Wyper},
  {DeVore}, \& {Lynch}}]{2021ApJ...907...41K}
{Kumar}, P., {Karpen}, J.~T., {Antiochos}, S.~K., {et~al.} 2021, \apj, 907, 41, \dodoi{10.3847/1538-4357/abca8b}

\bibitem[{{Kundu} {et~al.}(1995){Kundu}, {Raulin}, {Nitta}, {Hudson},
  {Shimojo}, {Shibata}, \& {Raoult}}]{1995ApJ...447L.135K}
{Kundu}, M.~R., {Raulin}, J.~P., {Nitta}, N., {et~al.} 1995, \apjl, 447, L135,
  \dodoi{10.1086/309567}

\bibitem[{{Lee} {et~al.}(2020){Lee}, {Karpen}, {Liu}, \&
  {Wang}}]{2020ApJ...893..158L}
{Lee}, J., {Karpen}, J.~T., {Liu}, C., \& {Wang}, H. 2020, \apj, 893, 158,
  \dodoi{10.3847/1538-4357/ab80c4}

\bibitem[{{Lemen} {et~al.}(2012){Lemen}, {Title}, {Akin}, {Boerner}, {Chou},
  {Drake}, {Duncan}, {Edwards}, {Friedlaender}, {Heyman}, {Hurlburt}, {Katz},
  {Kushner}, {Levay}, {Lindgren}, {Mathur}, {McFeaters}, {Mitchell}, {Rehse},
  {Schrijver}, {Springer}, {Stern}, {Tarbell}, {Wuelser}, {Wolfson}, {Yanari},
  {Bookbinder}, {Cheimets}, {Caldwell}, {Deluca}, {Gates}, {Golub}, {Park},
  {Podgorski}, {Bush}, {Scherrer}, {Gummin}, {Smith}, {Auker}, {Jerram},
  {Pool}, {Soufli}, {Windt}, {Beardsley}, {Clapp}, {Lang}, \&
  {Waltham}}]{2012SoPh..275...17L}
{Lemen}, J.~R., {Title}, A.~M., {Akin}, D.~J., {et~al.} 2012, \solphys, 275,
  17, \dodoi{10.1007/s11207-011-9776-8}

\bibitem[{{Li} \& {Yang}(2019)}]{2019ApJ...872...87L}
{Li}, H., \& {Yang}, J. 2019, \apj, 872, 87, \dodoi{10.3847/1538-4357/aafb3a}

\bibitem[{{Li} {et~al.}(2017){Li}, {Jiang}, {Yang}, {Qu}, {Yang}, {Xu}, {Bi},
  {Hong}, \& {Chen}}]{2017ApJ...842L..20L}
{Li}, H., {Jiang}, Y., {Yang}, J., {et~al.} 2017, \apjl, 842, L20,
  \dodoi{10.3847/2041-8213/aa762c}

\bibitem[{{Li} {et~al.}(2018){Li}, {Yang}, {Zhang}, {Hou}, \&
  {Zhang}}]{2018ApJ...859..122L}
{Li}, T., {Yang}, S., {Zhang}, Q., {Hou}, Y., \& {Zhang}, J. 2018, \apj, 859,
  122, \dodoi{10.3847/1538-4357/aabe84}

\bibitem[{{Liu} {et~al.}(2018){Liu}, {Erd{\'e}lyi}, {Wang}, \&
  {Liu}}]{2018ApJ...852...10L}
{Liu}, J., {Erd{\'e}lyi}, R., {Wang}, Y., \& {Liu}, R. 2018, \apj, 852, 10,
  \dodoi{10.3847/1538-4357/aa992d}

\bibitem[{{Liu} {et~al.}(2019){Liu}, {Wang}, \&
  {Erd{\'e}lyi}}]{2019FrASS...6...44L}
{Liu}, J., {Wang}, Y., \& {Erd{\'e}lyi}, R. 2019, Frontiers in Astronomy and
  Space Sciences, 6, 44, \dodoi{10.3389/fspas.2019.00044}

\bibitem[{{Liu} {et~al.}(2015){Liu}, {Wang}, {Shen}, {Liu}, {Pan}, \&
  {Wang}}]{2015ApJ...813..115L}
{Liu}, J., {Wang}, Y., {Shen}, C., {et~al.} 2015, \apj, 813, 115,
  \dodoi{10.1088/0004-637X/813/2/115}

\bibitem[{{Liu}(2008)}]{2008SoPh..249...75L}
{Liu}, Y. 2008, \solphys, 249, 75, \dodoi{10.1007/s11207-008-9176-x}

\bibitem[{{Liu} {et~al.}(2005){Liu}, {Su}, {Morimoto}, {Kurokawa}, \&
  {Shibata}}]{2005ApJ...628.1056L}
{Liu}, Y., {Su}, J.~T., {Morimoto}, T., {Kurokawa}, H., \& {Shibata}, K. 2005,
  \apj, 628, 1056, \dodoi{10.1086/431145}

\bibitem[{{Masson} {et~al.}(2009){Masson}, {Pariat}, {Aulanier}, \&
  {Schrijver}}]{2009ApJ...700..559M}
{Masson}, S., {Pariat}, E., {Aulanier}, G., \& {Schrijver}, C.~J. 2009, \apj,
  700, 559, \dodoi{10.1088/0004-637X/700/1/559}

\bibitem[{{Miao} {et~al.}(2018){Miao}, {Liu}, {Li}, {Shen}, {Yang}, {Elmhamdi},
  {Kordi}, \& {Abidin}}]{2018ApJ...869...39M}
{Miao}, Y., {Liu}, Y., {Li}, H.~B., {et~al.} 2018, \apj, 869, 39,
  \dodoi{10.3847/1538-4357/aaeac1}

\bibitem[{{Mishra} \& {Teriaca}(2023)}]{2023JApA...44...20M}
{Mishra}, W., \& {Teriaca}, L. 2023, Journal of Astrophysics and Astronomy, 44,
  20, \dodoi{10.1007/s12036-023-09910-6}

\bibitem[{{Moore} {et~al.}(2010){Moore}, {Cirtain}, {Sterling}, \&
  {Falconer}}]{2010ApJ...720..757M}
{Moore}, R.~L., {Cirtain}, J.~W., {Sterling}, A.~C., \& {Falconer}, D.~A. 2010,
  \apj, 720, 757, \dodoi{10.1088/0004-637X/720/1/757}

\bibitem[{{Mulay} {et~al.}(2016){Mulay}, {Tripathi}, {Del Zanna}, \&
  {Mason}}]{2016A&A...589A..79M}
{Mulay}, S.~M., {Tripathi}, D., {Del Zanna}, G., \& {Mason}, H. 2016, \aap,
  589, A79, \dodoi{10.1051/0004-6361/201527473}

\bibitem[{{Nistic{\`o}} {et~al.}(2009){Nistic{\`o}}, {Bothmer}, {Patsourakos},
  \& {Zimbardo}}]{2009SoPh..259...87N}
{Nistic{\`o}}, G., {Bothmer}, V., {Patsourakos}, S., \& {Zimbardo}, G. 2009,
  \solphys, 259, 87, \dodoi{10.1007/s11207-009-9424-8}

\bibitem[{{Panesar} {et~al.}(2016){Panesar}, {Sterling}, \&
  {Moore}}]{2016ApJ...822L..23P}
{Panesar}, N.~K., {Sterling}, A.~C., \& {Moore}, R.~L. 2016, \apjl, 822, L23,
  \dodoi{10.3847/2041-8205/822/2/L23}

\bibitem[{{Panesar} {et~al.}(2017){Panesar}, {Sterling}, \&
  {Moore}}]{2017ApJ...844..131P}
{Panesar}, N.~K., {Sterling}, A.~C., \& {Moore}, R.~L. 2017, \apj, 844, 131, 
\dodoi{10.3847/1538-4357/aa7b77}

\bibitem[{{Pant} {et~al.}(2021){Pant}, {Majumdar}, {Patel}, {Chauhan},
  {Banerjee}, \& {Gopalswamy}}]{2021FrASS...8...73P}
{Pant}, V., {Majumdar}, S., {Patel}, R., {et~al.} 2021, Frontiers in Astronomy
  and Space Sciences, 8, 73, \dodoi{10.3389/fspas.2021.634358}

\bibitem[{{Pariat} {et~al.}(2009){Pariat}, {Antiochos}, \&
  {DeVore}}]{2009ApJ...691...61P}
{Pariat}, E., {Antiochos}, S.~K., \& {DeVore}, C.~R. 2009, \apj, 691, 61,
  \dodoi{10.1088/0004-637X/691/1/61}

\bibitem[{{Pesnell} {et~al.}(2012){Pesnell}, {Thompson}, \&
  {Chamberlin}}]{2012SoPh..275....3P}
{Pesnell}, W.~D., {Thompson}, B.~J., \& {Chamberlin}, P.~C. 2012, \solphys,
  275, 3, \dodoi{10.1007/s11207-011-9841-3}

\bibitem[{{Pirjola}(2005)}]{2005AdSpR..36.2231P}
{Pirjola}, R. 2005, Advances in Space Research, 36, 2231,
  \dodoi{10.1016/j.asr.2003.04.074}

\bibitem[{{Raulin} {et~al.}(1996){Raulin}, {Kundu}, {Hudson}, {Nitta}, \&
  {Raoult}}]{1996A&A...306..299R}
{Raulin}, J.~P., {Kundu}, M.~R., {Hudson}, H.~S., {Nitta}, N., \& {Raoult}, A.
  1996, \aap, 306, 299

\bibitem[{{Reid}(2020)}]{2020FrASS...7...56R}
{Reid}, H. A.~S. 2020, Frontiers in Astronomy and Space Sciences, 7, 56,
  \dodoi{10.3389/fspas.2020.00056}

\bibitem[{{Schrijver} \& {De Rosa}(2003)}]{2003SoPh..212..165S}
{Schrijver}, C.~J., \& {De Rosa}, M.~L. 2003, \solphys, 212, 165,
  \dodoi{10.1023/A:1022908504100}

\bibitem[{{Shen}(2021)}]{2021RSPSA.47700217S}
{Shen}, Y. 2021, Proceedings of the Royal Society of London Series A, 477, 217,
  \dodoi{10.1098/rspa.2020.0217}

\bibitem[{{Shen} {et~al.}(2018){Shen}, {Liu}, {Liu}, {Su}, {Tang}, \&
  {Miao}}]{2018ApJ...861..105S}
{Shen}, Y., {Liu}, Y., {Liu}, Y.~D., {et~al.} 2018, \apj, 861, 105,
  \dodoi{10.3847/1538-4357/aac9be}

\bibitem[{{Shen} {et~al.}(2012{\natexlab{a}}){Shen}, {Liu}, \&
  {Su}}]{2012ApJ...750...12S}
{Shen}, Y., {Liu}, Y., \& {Su}, J. 2012{\natexlab{a}}, \apj, 750, 12,
  \dodoi{10.1088/0004-637X/750/1/12}

\bibitem[{{Shen} {et~al.}(2012{\natexlab{b}}){Shen}, {Liu}, {Su}, \&
  {Deng}}]{2012ApJ...745..164S}
{Shen}, Y., {Liu}, Y., {Su}, J., \& {Deng}, Y. 2012{\natexlab{b}}, \apj, 745,
  164, \dodoi{10.1088/0004-637X/745/2/164}

\bibitem[{{Shen} {et~al.}(2011){Shen}, {Liu}, {Su}, \&
  {Ibrahim}}]{2011ApJ...735L..43S}
{Shen}, Y., {Liu}, Y., {Su}, J., \& {Ibrahim}, A. 2011, \apjl, 735, L43,
  \dodoi{10.1088/2041-8205/735/2/L43}

\bibitem[{{Shen} {et~al.}(2017){Shen}, {Liu}, {Su}, {Qu}, \&
  {Tian}}]{2017ApJ...851...67S}
{Shen}, Y., {Liu}, Y.~D., {Su}, J., {Qu}, Z., \& {Tian}, Z. 2017, \apj, 851,
  67, \dodoi{10.3847/1538-4357/aa9a48}

\bibitem[{{Shen} {et~al.}(2019{\natexlab{a}}){Shen}, {Qu}, {Zhou}, {Duan},
  {Tang}, \& {Yuan}}]{2019ApJ...885L..11S}
{Shen}, Y., {Qu}, Z., {Zhou}, C., {et~al.} 2019{\natexlab{a}}, \apjl, 885, L11,
  \dodoi{10.3847/2041-8213/ab4cf3}

\bibitem[{{Shen} {et~al.}(2019{\natexlab{b}}){Shen}, {Qu}, {Yuan}, {Chen},
  {Duan}, {Zhou}, {Tang}, {Huang}, \& {Liu}}]{2019ApJ...883..104S}
{Shen}, Y., {Qu}, Z., {Yuan}, D., {et~al.} 2019{\natexlab{b}}, \apj, 883, 104,
  \dodoi{10.3847/1538-4357/ab3a4d}

\bibitem[{{Shibata} {et~al.}(1992){Shibata}, {Ishido}, {Acton}, {Strong},
  {Hirayama}, {Uchida}, {McAllister}, {Matsumoto}, {Tsuneta}, {Shimizu},
  {Hara}, {Sakurai}, {Ichimoto}, {Nishino}, \& {Ogawara}}]{1992PASJ...44L.173S}
{Shibata}, K., {Ishido}, Y., {Acton}, L.~W., {et~al.} 1992, \pasj, 44, L173

\bibitem[{{Shimojo} \& {Shibata}(2000)}]{2000ApJ...542.1100S}
{Shimojo}, M., \& {Shibata}, K. 2000, \apj, 542, 1100, \dodoi{10.1086/317024}

\bibitem[{{Solanki} {et~al.}(2019){Solanki}, {Srivastava}, {Rao}, \&
  {Dwivedi}}]{2019SoPh..294...68S}
{Solanki}, R., {Srivastava}, A.~K., {Rao}, Y.~K., \& {Dwivedi}, B.~N. 2019,
  \solphys, 294, 68, \dodoi{10.1007/s11207-019-1453-3}

\bibitem[{{Song} {et~al.}(2023){Song}, {Zhang}, {Li}, {Yang}, {Xia}, {Zheng},
  \& {Chen}}]{2023ApJ...942...19S}
{Song}, H., {Zhang}, J., {Li}, L., {et~al.} 2023, \apj, 942, 19,
  \dodoi{10.3847/1538-4357/aca6e0}

\bibitem[{{Sterling} {et~al.}(2015){Sterling}, {Moore}, {Falconer}, \&
  {Adams}}]{2015Natur.523..437S}
{Sterling}, A.~C., {Moore}, R.~L., {Falconer}, D.~A., \& {Adams}, M. 2015,
  \nat, 523, 437, \dodoi{10.1038/nature14556}

\bibitem[{{Sun} {et~al.}(2023){Sun}, {Li}, {Tian}, {Hou}, {Hou}, {Chen}, {Bai},
  \& {Deng}}]{2023ApJ...953..148S}
{Sun}, Z., {Li}, T., {Tian}, H., {et~al.} 2023, \apj, 953, 148,
  \dodoi{10.3847/1538-4357/ace5b1}

\bibitem[{{Tan} {et~al.}(2022){Tan}, {Shen}, {Zhou}, {Duan}, {Tang}, {Zhou}, \&
  {Yao}}]{2022MNRAS.516L..12T}
{Tan}, S., {Shen}, Y., {Zhou}, X., {et~al.} 2022, \mnras, 516, L12,
  \dodoi{10.1093/mnrasl/slac069}

\bibitem[{{Tan} {et~al.}(2023){Tan}, {Shen}, {Zhou}, {Tang}, {Zhou}, {Duan}, \&
  {Yao}}]{2023MNRAS.520.3080T}
{Tan}, S., {Shen}, Y., {Zhou}, X., {et~al.} 2023, \mnras, 520, 3080, \dodoi{10.1093/mnras/stad295}

\bibitem[{{Tang} {et~al.}(2021){Tang}, {Shen}, {Zhou}, {Duan}, {Zhou}, {Tan},
  \& {Elmhamdi}}]{2021ApJ...912L..15T}
{Tang}, Z., {Shen}, Y., {Zhou}, X., {et~al.} 2021, \apjl, 912, L15,
  \dodoi{10.3847/2041-8213/abf73a}

\bibitem[{{Thejappa} {et~al.}(2007){Thejappa}, {MacDowall}, \&
  {Kaiser}}]{2007ApJ...671..894T}
{Thejappa}, G., {MacDowall}, R.~J., \& {Kaiser}, M.~L. 2007, \apj, 671, 894,
  \dodoi{10.1086/522664}

\bibitem[{{Thompson}(2006)}]{2006A&A...449..791T}
{Thompson}, W.~T. 2006, \aap, 449, 791, \dodoi{10.1051/0004-6361:20054262}

\bibitem[{{Vourlidas} {et~al.}(2017){Vourlidas}, {Balmaceda}, {Stenborg}, \&
  {Dal Lago}}]{2017ApJ...838..141V}
{Vourlidas}, A., {Balmaceda}, L.~A., {Stenborg}, G., \& {Dal Lago}, A. 2017,
  \apj, 838, 141, \dodoi{10.3847/1538-4357/aa67f0}

\bibitem[{{Vourlidas} {et~al.}(2013){Vourlidas}, {Lynch}, {Howard}, \&
  {Li}}]{2013SoPh..284..179V}
{Vourlidas}, A., {Lynch}, B.~J., {Howard}, R.~A., \& {Li}, Y. 2013, \solphys,
  284, 179, \dodoi{10.1007/s11207-012-0084-8}

\bibitem[{{Wang} \& {Liu}(2012)}]{2012ApJ...760..101W}
{Wang}, H., \& {Liu}, C. 2012, \apj, 760, 101,
  \dodoi{10.1088/0004-637X/760/2/101}

\bibitem[{{Wang} {et~al.}(2007){Wang}, {Biersteker}, {Sheeley}, {Koutchmy},
  {Mouette}, \& {Druckm{\"u}ller}}]{2007ApJ...660..882W}
{Wang}, Y.~M., {Biersteker}, J.~B., {Sheeley}, N.~R., J., {et~al.} 2007, \apj,
  660, 882, \dodoi{10.1086/512480}

\bibitem[{{Wang} \& {Hess}(2023)}]{2023ApJ...952...85W}
{Wang}, Y.~M., \& {Hess}, P. 2023, \apj, 952, 85,
  \dodoi{10.3847/1538-4357/acd638}

\bibitem[{{Wang} {et~al.}(1998){Wang}, {Sheeley}, {Socker}, {Howard},
  {Brueckner}, {Michels}, {Moses}, {St. Cyr}, {Llebaria}, \&
  {Delaboudini{\`e}re}}]{1998ApJ...508..899W}
{Wang}, Y.~M., {Sheeley}, N.~R., J., {Socker}, D.~G., {et~al.} 1998, \apj, 508,
  899, \dodoi{10.1086/306450}

\bibitem[{{Wuelser} {et~al.}(2004){Wuelser}, {Lemen}, {Tarbell}, {Wolfson},
  {Cannon}, {Carpenter}, {Duncan}, {Gradwohl}, {Meyer}, {Moore}, {Navarro},
  {Pearson}, {Rossi}, {Springer}, {Howard}, {Moses}, {Newmark},
  {Delaboudiniere}, {Artzner}, {Auchere}, {Bougnet}, {Bouyries}, {Bridou},
  {Clotaire}, {Colas}, {Delmotte}, {Jerome}, {Lamare}, {Mercier}, {Mullot},
  {Ravet}, {Song}, {Bothmer}, \& {Deutsch}}]{2004SPIE.5171..111W}
{Wuelser}, J.-P., {Lemen}, J.~R., {Tarbell}, T.~D., {et~al.} 2004, in Society
  of Photo-Optical Instrumentation Engineers (SPIE) Conference Series, Vol.
  5171, Telescopes and Instrumentation for Solar Astrophysics, ed.
  S.~{Fineschi} \& M.~A. {Gummin}, 111--122, \dodoi{10.1117/12.506877}

\bibitem[{{Wyper} {et~al.}(2017){Wyper}, {Antiochos}, \&
  {DeVore}}]{2017Natur.544..452W}
{Wyper}, P.~F., {Antiochos}, S.~K., \& {DeVore}, C.~R. 2017, \nat, 544, 452,
  \dodoi{10.1038/nature22050}

\bibitem[{{Wyper} {et~al.}(2021){Wyper}, {Antiochos}, {DeVore}, {Lynch},
  {Karpen}, \& {Kumar}}]{2021ApJ...909...54W}
{Wyper}, P.~F., {Antiochos}, S.~K., {DeVore}, C.~R., {et~al.} 2021, \apj, 909,
  54, \dodoi{10.3847/1538-4357/abd9ca}

\bibitem[{{Wyper} {et~al.}(2018){Wyper}, {DeVore}, \&
  {Antiochos}}]{2018ApJ...852...98W}
{Wyper}, P.~F., {DeVore}, C.~R., \& {Antiochos}, S.~K. 2018, \apj, 852, 98,
  \dodoi{10.3847/1538-4357/aa9ffc}

\bibitem[{{Xing} {et~al.}(2018){Xing}, {Li}, {Jiang}, {Cheng}, \&
  {Ding}}]{2018ApJ...857L..14X}
{Xing}, C., {Li}, H.~C., {Jiang}, B., {Cheng}, X., \& {Ding}, M.~D. 2018,
  \apjl, 857, L14, \dodoi{10.3847/2041-8213/aabbb1}

\bibitem[{{Yang} {et~al.}(2020{\natexlab{a}}){Yang}, {Hong}, {Li}, \&
  {Jiang}}]{2020ApJ...900..158Y}
{Yang}, J., {Hong}, J., {Li}, H., \& {Jiang}, Y. 2020{\natexlab{a}}, \apj, 900,
  158, \dodoi{10.3847/1538-4357/aba7c0}

\bibitem[{{Yang} {et~al.}(2023){Yang}, {Yan}, {Xue}, {Chen}, {Wang}, {Xu}, \&
  {Li}}]{2023ApJ...945...96Y}
{Yang}, L., {Yan}, X., {Xue}, Z., {et~al.} 2023, \apj, 945, 96,
  \dodoi{10.3847/1538-4357/acb6f6}

\bibitem[{{Yang} {et~al.}(2024){Yang}, {Yan}, {Xue}, {Xu}, {Zhang}, {Hou},
  {Wang}, {Chen}, \& {Li}}]{2024MNRAS.528.1094Y}
{Yang}, L., {Yan}, X., {Xue}, Z., {et~al.} 2024, \mnras, 528, 1094, \dodoi{10.1093/mnras/stad3876}

\bibitem[{{Yang} {et~al.}(2020{\natexlab{b}}){Yang}, {Zhang}, {Xu}, {Zhang},
  {Zhong}, \& {Guo}}]{2020ApJ...898..101Y}
{Yang}, S., {Zhang}, Q., {Xu}, Z., {et~al.} 2020{\natexlab{b}}, \apj, 898, 101,
  \dodoi{10.3847/1538-4357/ab9ac7}

\bibitem[{{Zhang}(2024)}]{2024RvMPP...8....7Z}
{Zhang}, Q. 2024, Reviews of Modern Plasma Physics, 8, 7,
  \dodoi{10.1007/s41614-024-00144-9}

\bibitem[{{Zhang} {et~al.}(2021){Zhang}, {Huang}, {Hou}, {Li}, {Ning}, \&
  {Wu}}]{2021A&A...647A.113Z}
{Zhang}, Q.~M., {Huang}, Z.~H., {Hou}, Y.~J., {et~al.} 2021, \aap, 647, A113,
  \dodoi{10.1051/0004-6361/202038924}

\bibitem[{{Zhang} {et~al.}(2022{\natexlab{a}}){Zhang}, {Zhang}, {Dai}, {Li}, \&
  {Ji}}]{2022SoPh..297..138Z}
{Zhang}, Y., {Zhang}, Q., {Dai}, J., {Li}, D., \& {Ji}, H. 2022{\natexlab{a}},
  \solphys, 297, 138, \dodoi{10.1007/s11207-022-02072-8}

\bibitem[{{Zhang} {et~al.}(2022{\natexlab{b}}){Zhang}, {Zhang}, {Song}, {Li},
  {Dai}, {Xu}, \& {Ji}}]{2022ApJS..260...19Z}
{Zhang}, Y., {Zhang}, Q., {Song}, D., {et~al.} 2022{\natexlab{b}}, \apjs, 260,
  19, \dodoi{10.3847/1538-4365/ac5f4c}

\bibitem[{{Zheng} {et~al.}(2016){Zheng}, {Chen}, {Du}, \&
  {Li}}]{2016ApJ...819L..18Z}
{Zheng}, R., {Chen}, Y., {Du}, G., \& {Li}, C. 2016, \apjl, 819, L18,
  \dodoi{10.3847/2041-8205/819/2/L18}

\bibitem[{{Zhou} {et~al.}(2021{\natexlab{a}}){Zhou}, {Shen}, {Zhou}, {Tang},
  {Duan}, \& {Tan}}]{2021ApJ...923...45Z}
{Zhou}, C., {Shen}, Y., {Zhou}, X., {et~al.} 2021{\natexlab{a}}, \apj, 923, 45,
  \dodoi{10.3847/1538-4357/ac28a0}

\bibitem[{{Zhou} {et~al.}(2021{\natexlab{b}}){Zhou}, {Xia}, \&
  {Shen}}]{2021A&A...647A.112Z}
{Zhou}, C., {Xia}, C., \& {Shen}, Y. 2021{\natexlab{b}}, \aap, 647, A112,
  \dodoi{10.1051/0004-6361/202039558}

\bibitem[{{Zhou} {et~al.}(2017){Zhou}, {Zhang}, {Wang}, {Liu}, \&
  {Chintzoglou}}]{2017ApJ...851..133Z}
{Zhou}, Z., {Zhang}, J., {Wang}, Y., {Liu}, R., \& {Chintzoglou}, G. 2017,
  \apj, 851, 133, \dodoi{10.3847/1538-4357/aa9bd9}

\end{thebibliography}
\end{document}